\documentclass[aps,prb,reprint,superscriptaddress,amsmath,amssymb]{revtex4-2}
\usepackage{graphicx}
\usepackage{bm}
\usepackage{xcolor}
\usepackage{xr-hyper}
\usepackage[colorlinks=true,allcolors=blue!50!black]{hyperref}
\newcommand{\km}{\kappa_m}
\newcommand{\dR}{\delta R}
\newcommand{\srot}{s_{\rm rot}}
\newcommand{\seig}{s_{\rm eig}}
\newcommand{\WmK}{W/(m\,K)}

\begin{document}

\title{Where heat flows and where to change it: gauge-free sensitivity maps of thermal conductivity in disordered solids}

\author{D. A. Drabold}
\email{drabold@ohio.edu}
\affiliation{Department of Physics and Astronomy, Ohio University, Athens, Ohio 45701, USA}
\author{C. Ugwumadu}
\affiliation{Theoretical Division (T-4), Los Alamos National Laboratory, Los Alamos, New Mexico 87545, USA}

\date{\today}

\begin{abstract}
Designing materials for thermal management requires knowing where heat flows, not only how much.
Within harmonic (Allen--Feldman) theory we develop two local descriptions and test them on amorphous silicon.
The first, a projection of the conductivity $\kappa$ onto atoms, depends on how the potential energy is
divided among atoms (a ``gauge''); we find the dependence moderate, 23\% of the map's spatial standard
deviation at single sites and 9.5\% over regions of radius 4~\AA, apart from a uniform offset of 0.8\% of $\kappa$. The second avoids the division altogether. It
is the exact response of $\kappa$ to the mass of each atom and the stiffness of each bond, it obeys an Euler sum
rule, and it contains the heat-flow map as one of three exact terms. In amorphous and paracrystalline silicon, with ACE,
GAP-18 and Keating force constants, the two maps are anti-correlated: where little heat flows a heavier atom
lowers $\kappa$, and where much heat flows it tends to raise it. The explanation is local softness. The
harmonic vibrational amplitude of each atom predicts both maps, and at room temperature the
anti-correlation largely disappears at fixed amplitude, in a 32\,768-atom replica as well; below about
100~K a component remains that softness does not capture. Thus soft regions, which carry little heat, act
through the mixing of vibrational modes about 1--2~meV apart. In paracrystalline silicon the grains conduct more heat per atom,
while the disordered matrix controls the response. Exact recalculations show that the response map guides
the placement of dispersed substituents and locates the region that limits transport across an
amorphous/crystalline interface.
\end{abstract}

\maketitle

\section{Introduction}

Heat conduction in solids, especially disordered ones, is a fundamental problem for basic science and
for applications. Heat must be removed from nanoscale devices through amorphous dielectrics and across
interfaces~\cite{Pop2010energy,Cahill2014nanoscale}; thermoelectrics need a low lattice
conductivity~\cite{Snyder2008complex}; phase-change memories and thermal barrier coatings depend on the
conductivity of disordered phases~\cite{Wuttig2007phase,Padture2002thermal}. The theory is well developed.
Crystals are treated by Boltzmann transport with first-principles phonons~\cite{Broido2007intrinsic};
glasses by the harmonic theory of Allen and Feldman (AF) and its unified extensions that include
anharmonic damping~\cite{Simoncelli2019unified,Isaeva2019unified}; and any of them by Green--Kubo molecular
dynamics. Each delivers a single number, the conductivity $\kappa$. Machine-learned interatomic potentials
now give force constants of near first-principles quality for $10^4$--$10^6$ atoms~\cite{Deringer2019machine},
so local descriptions of heat transport are within reach, and they are what design needs: where in a
material does heat flow, and where would a change matter? We use amorphous silicon as a proving ground.
It is the simplest tetrahedral network, its structure and vibrations are well characterized, and accurate
potentials exist. The methods apply to any harmonic network.

Amorphous silicon (a-Si) conducts heat poorly, and it is worth knowing both why and where. The
AF theory answers the first
question~\cite{Allen1989thermal,Allen1993disordered,Feldman1993thermal,Allen1999diffusons}: heat is carried
mostly by \emph{diffusons}, delocalized but non-propagating vibrations that pass energy between modes of
nearly equal frequency. Long-wavelength propagons add a contribution that depends on sample size, and
localized vibrations (locons) carry almost nothing; the measured conductivity of a-Si films is about
1--3~\WmK{}, depending on thickness~\cite{Braun2016size,Regner2013broadband}. The second question, where in
the network heat is carried and where a local change would alter $\kappa$, is harder. It is the subject
of this paper.

In earlier work we projected $\kappa$ onto atoms, the site-projected thermal conductivity (SPTC)
$\zeta({\ell})$~\cite{Gautam2025siteprojected,Ugwumadu2026mapping,Gautam2026lattice,Ugwumadu2026seamless}, and applied it to the phase-change material Ge$_2$Sb$_2$Te$_5$~\cite{Nepal2026electronic}. The projection has a
feature familiar to anyone who has computed a Mulliken charge: it is not unique. The heat current depends
on how the potential energy is divided among atoms, and this division (a ``gauge'') is
arbitrary~\cite{Ercole2016gauge,Marcolongo2016microscopic}; for machine-learned potentials the atomic
energies are themselves a by-product of the fit~\cite{Chong2023robustness}. The total $\kappa$ is
indifferent to the gauge; the site values are \emph{not}. Thus a reader is entitled to ask whether a map
of SPTC describes the network or the bookkeeping.

We answer in two ways. First, we measure the gauge dependence: it is moderate at single sites and small
over regions a few \AA\ across. Second, and more interesting, we ask a question with no gauge ambiguity at
all: if one changes the mass of one atom, or the stiffness of one bond, how does $\kappa$ change? Since
the AF expression for $\kappa$ involves no division of the energy among atoms, neither do its derivatives
(Sec.~\ref{sec:theory}). We derive the exact gradient of the AF conductivity
with respect to the dynamical matrix, contract it into mass and bond sensitivity maps, and decompose the
mass map exactly into three terms, one of which is the SPTC itself. The answer is at first surprising. In the regions that carry the \emph{least} heat a heavier atom lowers
$\kappa$, and in those that carry the most it tends to raise it: the \emph{sign} of the response is
anti-correlated with the heat flow, and averaged over a few \AA\ its magnitude is larger where little heat
flows. A single local quantity, the harmonic vibrational amplitude $\dR({\ell})$ of each atom, mediates the
relation, and at fixed $\dR$ the anti-correlation between the two maps largely vanishes. Soft,
large-amplitude sites (``rattlers'') contribute little to transport, but a change there mixes the
vibrational modes that do.

Soft regions are no strangers to the physics of glasses. Quasi-localized low-frequency
modes identify ``soft spots'' that predict where plastic rearrangements occur under
shear~\cite{Manning2011soft,Schober1991localized,Xu2010anharmonic}, and elastic heterogeneity has been linked to vibrational
anomalies and to the thermal conductivity of glasses~\cite{Mizuno2013elastic,Mizuno2016elastic}. Our results give the
thermal counterpart: in a-Si the soft regions are both where heat flows least and where its flow can be
changed most.

The paper is organized as follows. Section~\ref{sec:theory} defines the SPTC and the sensitivity maps,
derives the exact gradient and its decomposition, and shows that $\dR$ is a local diagonal of the
pseudo-inverse of the dynamical matrix. Section~\ref{sec:models} describes the models and calculations.
Section~\ref{sec:results} establishes the robustness of $\zeta$, the anti-correlation of the two maps and
its mechanism, its temperature dependence, paracrystalline silicon, consequences for design, and an
application to an
amorphous/crystalline interface. Section~\ref{sec:discussion} discusses limitations and connections to
other work.

\section{Theory}\label{sec:theory}

\subsection{Allen--Feldman conductivity and its site projection}

Let $\Phi$ be the force-constant matrix, $m_{\ell}$ the mass of atom ${\ell}$ (atoms are labelled ${\ell},{\ell'}$
throughout, as in Hardy's notation~\cite{Hardy1963energy}), and
$D=M^{-1/2}\Phi M^{-1/2}$ the dynamical matrix, with eigenpairs $De_i=\lambda_ie_i$,
$\lambda_i=\omega_i^2$. Cartesian indices $a,b$ run over the $3N$ degrees of freedom and ${\ell}(a)$ is
the atom of $a$. With the bond-weighted operators $\Gamma^\alpha=D\circ R^\alpha$, where $\circ$ denotes the
elementwise (Hadamard) product, $(D\circ R^\alpha)_{ab}=D_{ab}R^\alpha_{ab}$,
$R^\alpha_{ab}=(R_{{\ell}(b)}-R_{{\ell}(a)})^\alpha$ (minimum image), the AF conductivity is
\begin{equation}
\kappa=\sum_\alpha\sum_{i\neq j}F(\lambda_i,\lambda_j)\big(P^\alpha_{ij}\big)^2,\qquad
P^\alpha=E^{\mathsf T}\Gamma^\alpha E,
\label{eq:kappa}
\end{equation}
where $E$ holds the eigenvectors and~\cite{Allen1993disordered}
\begin{align}
F(\lambda_i,\lambda_j)&=\frac{\pi}{12V}\,\frac12\Big(\frac{C_i}{\omega_i^2}+\frac{C_j}{\omega_j^2}\Big)
\frac{(\omega_i+\omega_j)^2}{4\omega_i\omega_j}\,L_\eta(\omega_i-\omega_j),\label{eq:F}\\
L_\eta(x)&=\frac{\eta/\pi}{x^2+\eta^2},\nonumber
\end{align}
with $V$ the volume, $C_i$ the heat capacity of mode $i$ (quantum unless stated) and $\eta$ the half-width of
the Lorentzian; with per-mode widths $\gamma_i$, $\eta\to(\gamma_i+\gamma_j)/2$. In SI units ($D$ in s$^{-2}$, $R$ in m,
$C$ in J/K) $\kappa$ is in \WmK{}. Modes below 0.5~meV, which include the three translations, are excluded. We quote $\eta$ in kelvin,
as the temperature $k_B\eta_K=\hbar\eta$; $\eta=2$~K corresponds to $\hbar\eta=0.17$~meV. Writing $\kappa$ as a trace over sites defines the site projection
\begin{equation}
\zeta({\ell})=\sum_\alpha\sum_{a\in {\ell}}\sum_b \Gamma^\alpha_{ab}\big[E(F\circ P^\alpha)E^{\mathsf T}\big]_{ab},
\qquad \sum_{\ell}\zeta({\ell})=\kappa .
\label{eq:zeta}
\end{equation}
Anchoring each bond current at the row atom is one choice of local energy, the ``AF gauge''. Any other
choice, in practice the site energies of the potential, gives a different $\zeta$ with the same
sum~\cite{Ercole2016gauge}. The weight $F$ is symmetric in $(i,j)$ through the factor
$\tfrac12(C_i/\omega_i^2+C_j/\omega_j^2)$; this leaves $\kappa$ unchanged and makes $\zeta$ independent of
whether a bond current is anchored at its row or its column atom (Appendix~\ref{app:remarks}).

\subsection{Gauge freedom}\label{sec:gauge}

A local energy assigns to each atom ${\ell}$ a quadratic form $h_{\ell}=p_{\ell}^2/2m_{\ell}+\tfrac12\,u^{\mathsf T}H^{({\ell})}u$
with $\sum_{\ell}H^{({\ell})}=\Phi$. The total energy fixes $\Phi$ but not the $H^{({\ell})}$: this is the gauge freedom.
The AF choice gives each atom its diagonal block and half of each mixed block; a many-body potential
with site energies $\varepsilon_{\ell}$ gives $H^{({\ell})}_{mn}=\partial^2\varepsilon_{\ell}/\partial u_m\,\partial u_n$, which
couples all atoms within the range $r_s$ of $\varepsilon_{\ell}$. Energy conservation gives the site currents
$\mathbf j_{\ell}=\sum_{mn}u_m\mathbf A^{({\ell})}_{mn}\dot u_n$ with
$\mathbf A^{({\ell})}_{mn}=(\mathbf R_{\ell}-\mathbf R_n)H^{({\ell})}_{mn}$~\cite{Hardy1963energy}, and in the mode
representation the current matrix element is proportional to
\begin{equation}
P^{\rm eff}_{ij}=-2B^{\rm a}_{ij}+2B^{\rm s}_{ij}\,\frac{\omega_i-\omega_j}{\omega_i+\omega_j},
\label{eq:Peff}
\end{equation}
where $B^{\rm a}$ and $B^{\rm s}$ are the antisymmetric and symmetric parts of
$B=E^{\mathsf T}\tilde{\mathbf A}E$, with $\tilde{\mathbf A}=\sum_{\ell}\mathbf A^{({\ell})}$ mass weighted
(Supplemental Material, Sec.~\textcolor{red}{S8}). In the AF gauge $B$ is antisymmetric and $P^{\rm eff}=P$. Three results
follow.

\emph{(i) $\kappa$ is gauge invariant up to $O(\eta)$.} Under $H^{({\ell})}\to H^{({\ell})}+\Delta H^{({\ell})}$ with
$\sum_{\ell}\Delta H^{({\ell})}=0$, the antisymmetric part of the total current changes by
$\tfrac12(\mathbf R_m-\mathbf R_n)\sum_{\ell}\Delta H^{({\ell})}_{mn}=0$. Only $B^{\rm s}$ depends on the gauge, and in
Eq.~(\ref{eq:Peff}) it is multiplied by $(\omega_i-\omega_j)/(\omega_i+\omega_j)$, which vanishes where the
Lorentzian concentrates the weight. For Stillinger--Weber, where every gauge can be computed exactly,
the gauge differences of $\kappa$ for a 196-atom model (Sec.~\ref{sec:models}) are 0.29, 0.50, 0.95 and 1.9\% at $\eta=1$, 2, 4 and 8~K,
linear in $\eta$.

\emph{(ii) Site values are not invariant.} For an individual site the antisymmetric part changes by
$\tfrac12(\mathbf R_m-\mathbf R_n)\Delta H^{({\ell})}_{mn}\neq0$. This piece sums to zero over sites, but it is not
the time derivative of a bounded quantity and carries no factor $(\omega_i-\omega_j)/(\omega_i+\omega_j)$, so the
argument that protects $\kappa$ does not protect $\zeta({\ell})$: the single-site uncertainty remains as
$\eta\to0$. For Stillinger--Weber it is 16, 13, 11 and 12\% of the spread of $\zeta$ at $\eta=1$, 2, 4 and 8~K.

\begin{figure*}[!t]
\includegraphics[width=\textwidth]{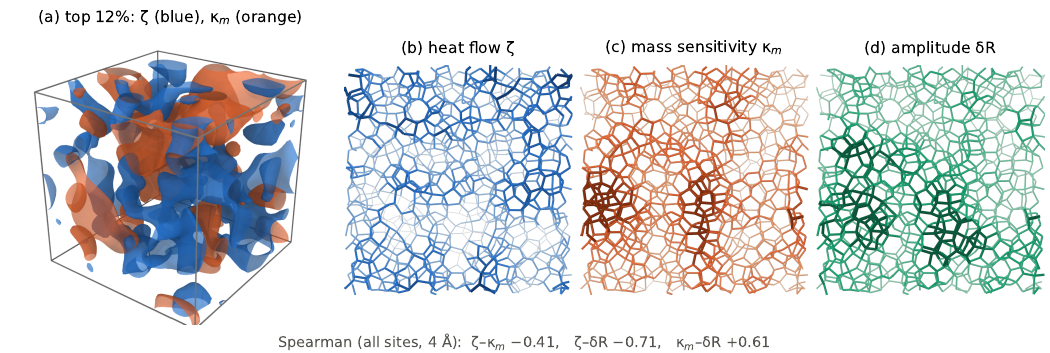}
\caption{The 4096-atom a-Si network (ACE, $\eta=2$~K). (a) The whole cell: translucent surfaces enclose the 12\%
of the volume where the heat-flow map $\zeta$ (blue) and the mass-sensitivity map $\km$ (orange), smoothed with
a Gaussian of 2.2~\AA, are largest. The two regions share 0.27\% of the volume, against 1.44\% for independent
maps. (b)--(d) A 9~\AA\ slab, with bonds coloured and weighted by fields averaged over 4~\AA\ spheres: (b) the
heat-flow map $\zeta$, (c) the gauge-free mass-sensitivity map $\km$, and (d) the harmonic vibrational amplitude
$\dR$ (300~K). Darker colours and thicker bonds indicate larger field values. In the soft, large-amplitude
regions, which carry little heat, $\km$ is positive and large: a heavier atom there lowers $\kappa$.}
\label{fig:maps}
\end{figure*}

\begin{figure*}[!t]
\includegraphics[width=.85\textwidth]{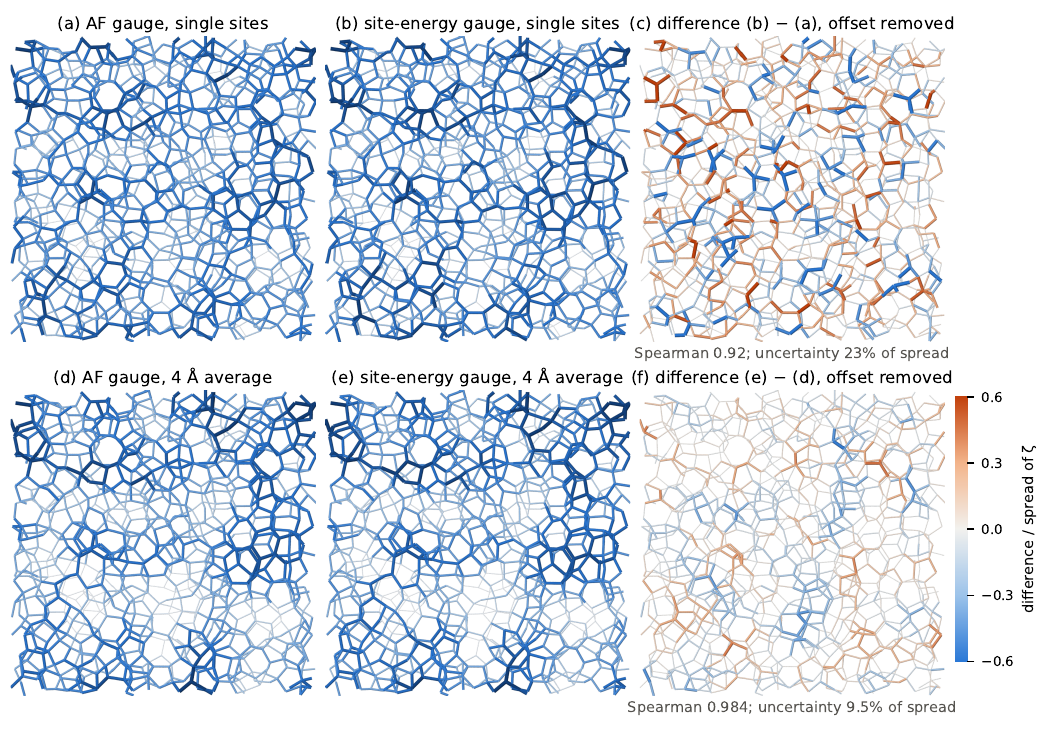}
\caption{The gauge question at a glance: the heat-flow map $\zeta$ of the slab of Fig.~\ref{fig:maps} at single sites (top row) and
averaged over 4~\AA\ spheres (bottom row). (a), (d) AF gauge; (b), (e) site-energy gauge of the ACE potential, on one colour
scale per row; (c), (f) their difference, in units of the spatial standard deviation of $\zeta$ in that row, on one diverging
scale for both rows. The two gauges look alike at a glance because their difference is small compared with the range of $\zeta$; the difference maps show
where they part, after removal of the uniform offset between the gauges (0.8\% of $\kappa$). Between the gauges, Spearman 0.92 and gauge uncertainty 23\% of the spatial spread at single sites, 0.984 and 9.5\% at
4~\AA. At 45\% of single sites the gauges differ by more than 0.3 of the spread, against 11\% of 4~\AA\ regions.}
\label{fig:gauge}
\end{figure*}

\emph{(iii) Regional values are protected up to a surface term.} Summed over a region $\Omega$, the change
involves $\sum_{{\ell}\in\Omega}\Delta H^{({\ell})}_{mn}$, which vanishes by $\sum_{\ell}\Delta H^{({\ell})}=0$ whenever atoms $m$
and $n$ lie deeper than $r_s$ inside $\Omega$. Only a shell of thickness $\sim r_s$ contributes, so the
local gauge ambiguity of a regional conductivity should fall roughly as $r_s/R$ for a region of size $R$. It does, roughly, with a small floor: for si4096
(ACE, $r_s=6.4$~\AA) it is 23\% of the spatial spread of $\zeta$ at single sites, 9.5\% at $R=4$~\AA\ and 4.6\% at
10~\AA, falling roughly as $r_s/R$ beyond about 4~\AA, as expected for a surface term (Supplemental Material,
Sec.~\textcolor{red}{S9}). A uniform $O(\eta)$ offset of 0.8\% of $\kappa$, the global difference of result (i), remains; we
quote it separately. The ranking of sites is nearly gauge-invariant beyond a few \AA\ (Spearman 0.984 at 4~\AA, 0.995 at
10~\AA), and for a map of where heat flows that is the claim that matters.
Thus the site projection is a well-defined description of where heat flows at the scale of a few
\AA{}, but not of individual atoms (Figs.~\ref{fig:maps} and \ref{fig:gauge}). The gauge-invariant generalization of a local conductivity is the
wave-vector-dependent $\kappa(q)$, whose gauge dependence vanishes as $q\to0$~\cite{Ercole2016gauge}.

The energy density of density-functional theory has the same ambiguity, and unique atomic energies follow from integrating over volumes on whose boundaries the gauge terms vanish~\cite{Chetty1992first,Yu2011energy}; result (iii) is an approximate analogue for the heat current.

\subsection{Gauge-free sensitivities and an Euler sum rule}

The site energies do not enter Eq.~(\ref{eq:kappa}), which is a function of the dynamical matrix and the
positions alone. Thus its derivatives with respect to masses and force constants are unique, and they
require \emph{no} partition of $\kappa$ among atoms; this is the sense in which we call them gauge free. At finite $\eta$, Eq.~(\ref{eq:kappa}) is the conductivity of the AF gauge, and a
conductivity built from another gauge differs from it by $O(\eta)$ [result (i) above]; the derivatives of such
a conductivity would differ from ours by the derivative of that difference, and all of them coincide as
$\eta\to0$. How large is the difference in practice? For Stillinger--Weber, where every gauge can be computed
exactly, we obtained the mass maps of the AF gauge and of two site-energy gauges by finite
differences in the 196-atom model of result (i): they differ by 0.1, 0.1, 0.2 and 0.4\% of the spread of
$\km$ at $\eta=1$, 2, 4 and 8~K, growing with $\eta$ and smaller than the gauge differences of $\kappa$ in the same
cell (0.3--1.9\%), and they rank the sites identically (Spearman rank correlation 1.000; the statistics are described in
Sec.~\ref{sec:models}). Thus, at least for this potential, the maps are gauge free for practical purposes at
finite $\eta$ as well.

We define the mass sensitivity
\begin{equation}
s_m({\ell})=-m_{\ell}\frac{\partial\kappa}{\partial m_{\ell}},\qquad \km({\ell})\equiv 2s_m({\ell}),
\end{equation}
where, since frequencies scale as $m^{-1/2}$, $\km({\ell})=\partial\kappa/\partial\ln m_{\ell}^{-1/2}$ is the response of $\kappa$ to the
local frequency scale at atom ${\ell}$; and the bond sensitivity $s_b({\ell},{\ell'})=d\kappa/d\epsilon|_{\epsilon=0}$, where $\epsilon$ is a small dimensionless parameter, the fractional change of the force constants of the pair: the blocks $\Phi_{{\ell}{\ell'}},\Phi_{{\ell'}{\ell}}$ are scaled by
$1+\epsilon$, with the self blocks compensated by $-\epsilon\,{\rm sym}(\Phi_{{\ell}{\ell'}})$, where
${\rm sym}(A)=\tfrac12(A+A^{\mathsf T})$, so that the perturbed matrix remains symmetric. For pair potentials $\Phi_{{\ell}{\ell'}}$ is symmetric and
this compensation preserves the acoustic sum rule exactly. It is well to observe that for many-body potentials
$\Phi_{{\ell}{\ell'}}$ is \emph{not} symmetric, and the row sum of atom ${\ell}$ changes by $\epsilon\,{\rm asym}(\Phi_{{\ell}{\ell'}})$, with
${\rm asym}(A)=\tfrac12(A-A^{\mathsf T})$ (and that of ${\ell'}$ by the opposite amount). No change of a single
non-symmetric block keeps both the symmetry of $D$ and the sum rule exactly. Fortunately the residual is
small: for
nearest-neighbour bonds ${\rm asym}(\Phi_{{\ell}{\ell'}})$ is 1--2\% of the block (median; 3--4\% at the 90th
percentile) for ACE and 4\% for Keating. Moreover, it cancels exactly when summed over all bonds of an atom, because
$\sum_{\ell'}\Phi_{{\ell}{\ell'}}=-\Phi_{{\ell}{\ell}}$ is symmetric, so the bond sensitivities, summed over all pairs $({\ell},{\ell'})$ within the range of $\Phi$, obey the
Euler rule below. The
alternative that scales only ${\rm sym}(\Phi_{{\ell}{\ell'}})$, which preserves both properties for each bond, gives
bond maps with Spearman correlation 0.986--0.998 with ours for nearest-neighbour bonds (sign changes for
0.3--2.9\% of them), but violates the Euler rule by 4--22\% (Supplemental Material, Table~\textcolor{red}{S1}).

Scaling all masses by $\mu$, together with $\eta\to\eta/\sqrt\mu$ and $T\to T/\sqrt\mu$, maps every
frequency $\omega\to\omega/\sqrt\mu$ and gives $\kappa\to\mu^{-1/2}\kappa$. Euler's theorem for this
homogeneous function yields (Appendix~\ref{app:euler})
\begin{equation}
\kappa=\sum_{\ell}\km({\ell})+\eta\frac{\partial\kappa}{\partial\eta}+T\Big(\frac{\partial\kappa}{\partial T}\Big)_{C}.
\label{eq:euler}
\end{equation}
Here the temperature derivative acts only through the heat capacities. Equation~(\ref{eq:euler}) checks one number, the sum of
the gradient over all sites. Thus it tests the current and eigenvalue parts and the $\eta$ and $T$
derivatives together, but it is
\emph{blind} to the spatial distribution: the rotation term of the
decomposition derived in Sec.~\ref{sec:grad} sums to zero identically. The site-by-site test is the comparison with finite
differences (Supplemental Material, Sec.~\textcolor{red}{S3}). The rule is also a consistency identity, not a partition: individual
$\km({\ell})$ are of both signs and typically larger than $\zeta({\ell})$ (for si4096 at $\eta=2$~K the median of
$|\km|$ is 1.8 times the mean of $\zeta$, and the standard deviation of $\km$ is 3.3 times it), and their sum reproduces $\kappa$
through large cancellations. We read $\km$ as a response. The same rule holds for the bond sensitivities summed over all pairs within the range of $\Phi$, not only
nearest-neighbour bonds.

\subsection{Exact gradient and decomposition}\label{sec:grad}

The gradient $G=\partial\kappa/\partial D$ has an explicit part from $\Gamma^\alpha=D\circ R^\alpha$ and a
spectral part from the eigenpairs. The spectral part is first-order perturbation theory in which divided
differences handle degeneracies, the Daleckii--Krein formula for functions of Hermitian
matrices~\cite{Daleckii1965integration,Bhatia1997matrix,Higham2008functions}; the bilinear structure of Eq.~(\ref{eq:kappa}) is
a double operator integral in the sense of Birman and Solomyak~\cite{Birman2003double}. We find
(Appendix~\ref{app:gradient})
\begin{align}
G&=E\tilde GE^{\mathsf T}+2\sum_\alpha R^\alpha\circ\big[E(F\circ P^\alpha)E^{\mathsf T}\big],
\label{eq:grad}\\
\tilde G_{ik}&=\frac{N_{ik}}{\lambda_i-\lambda_k},\qquad
N_{ik}=2\sum_\alpha\sum_jP^\alpha_{ij}P^\alpha_{kj}\big(F_{ij}-F_{kj}\big),
\label{eq:Gt}
\end{align}
for $i\neq k$. On the diagonal
$\tilde G_{ii}=\partial\kappa/\partial\lambda_i=2\sum_\alpha\sum_j(P^\alpha_{ij})^2\,\partial_1F(\lambda_i,\lambda_j)$ is the
partial derivative at fixed eigenvectors, which acts on $F$ only. For near-degenerate pairs
($|\lambda_i-\lambda_k|$ below a tolerance $10^{-7}$ relative to the spectral width)
$\tilde G_{ik}\to2\sum_\alpha\sum_jP^\alpha_{ij}P^\alpha_{kj}\,\partial_1F(\bar\lambda,\lambda_j)$; results
are insensitive to the tolerance. Both terms of Eq.~(\ref{eq:grad}) are symmetric matrices. In the first,
$\tilde G$ is symmetric because $N_{ik}$ and $\lambda_i-\lambda_k$ both change sign when $i$ and $k$ are
exchanged. In the second, $R^\alpha$ and $E(F\circ P^\alpha)E^{\mathsf T}$ are both antisymmetric, and the
elementwise product of two antisymmetric matrices is symmetric. This is as it must be: $D$ is symmetric, so
any change $\delta D$ is symmetric too, and the first-order change
$\delta\kappa=\sum_{ab}G_{ab}\,\delta D_{ab}$ involves only the symmetric part of $G$.

Since $\partial D_{ab}/\partial\ln m_{\ell}=-\tfrac12(\delta_{{\ell}(a),{\ell}}+\delta_{{\ell}(b),{\ell}})D_{ab}$, the mass
sensitivity is $s_m({\ell})=\sum_{a\in {\ell},b}G_{ab}D_{ab}$. Inserting Eq.~(\ref{eq:grad}) gives an exact
three-term decomposition,
\begin{equation}
\km({\ell})=4\zeta({\ell})+2\seig({\ell})+2\srot({\ell}),
\label{eq:decomp}
\end{equation}
\begin{equation}
\seig({\ell})=\sum_i\frac{\partial\kappa}{\partial\lambda_i}\lambda_ip_i({\ell}),\quad
\srot({\ell})=\sum_{i<k}N_{ik}\frac{\lambda_i+\lambda_k}{\lambda_i-\lambda_k}c_{ik}({\ell}),
\label{eq:terms}
\end{equation}
where $p_i({\ell})=\sum_{a\in {\ell}}E_{ai}^2$ is the weight of mode $i$ on atom ${\ell}$ and
$c_{ik}({\ell})=\sum_{a\in {\ell}}E_{ai}E_{ak}$ the local overlap of two modes. The first term arises because the
explicit part of the gradient contracted with $D$ is exactly twice the AF-gauge $\zeta$. The
eigenvalue-shift term $\seig$ is a sensitivity-weighted local density of states. The rotation term
$\srot$ sums to zero over sites, because $\sum_{\ell}c_{ik}({\ell})=\delta_{ik}$: it only redistributes
sensitivity in space.

\subsection{Local softness}\label{sec:dR}

The classical mean-square displacement of atom ${\ell}$ in the harmonic approximation is
\begin{equation}
\dR_{\ell}^2=\frac{k_BT}{m_{\ell}}\sum_i\frac{p_i({\ell})}{\lambda_i}
=\frac{k_BT}{m_{\ell}}\sum_{a\in {\ell}}\big[D^+\big]_{aa},
\label{eq:dR}
\end{equation}
where the sum excludes the three translations and $D^+$ is the pseudo-inverse of $D$
(Supplemental Material, Sec.~\textcolor{red}{S7}). It is the quantity used by Gautam~\emph{et al.} to identify
``rattling'' sites~\cite{Gautam2026lattice}. Two features matter here. First, $\dR_{\ell}^2$ is a local
spectral moment of the same mode weights $p_i({\ell})$ that enter $\seig$, weighted toward low frequencies.
Second, the right-hand form needs no eigenvectors: it can be evaluated with three sparse linear solves
per atom, at any system size; a complete map costs $3N$ solves. Since $k_BT/m_{\ell}$ is the same for every site of a monatomic network, $\dR$ enters our rank statistics
only through the local compliance $\sum_{a\in {\ell}}[D^+]_{aa}$, a structural property of the harmonic
network; its ranking, and every correlation built from it, is the same at all temperatures, and we use the
classical form throughout, quoting magnitudes at 300~K. The quantum amplitude, which tends to a $1/\omega$
weighting at low temperature, is examined in Sec.~\ref{sec:temperature}. We also use the local low-frequency weight $w_{\rm low}({\ell})=\sum_{\omega_i<12.4\,{\rm meV}}p_i({\ell})$.

\section{Models and calculations}\label{sec:models}

\paragraph*{Structures.}
The reference model is the 4096-atom continuous random network of Djordjevi\'c, Thorpe and
Wooten~\cite{Djordjevic1995computer} scaled to Si (si4096; cubic cell of 43.42~\AA, density 2.334~g/cm$^3$), relaxed in its fixed cell with each potential. After relaxation with the ACE potential
it has 5 three-fold and 23 five-fold atoms (0.68\% coordination defects at a 2.8~\AA\ cutoff). For
validation we use a 512-atom model of the same construction~\cite{Djordjevic1995computer} (si512, fully four-fold). Four independent networks test
generality: the 10\,000-atom Barkema--Mousseau model~\cite{Barkema2000highquality} (BM10k), relaxed with ACE at
zero pressure (0.56\% defects) and annealed at 800 and 1000~K and relaxed again (1.24\% and 1.51\%);
and three 1000-atom paracrystalline models R1--R3 of Rosset~\emph{et al.}~\cite{Rosset2025paracrystallinity}.
For the interface application we use the 2022-atom a-Si/c-Si model of Bernstein, Aziz and
Kaxiras~\cite{Bernstein1998interface}, relaxed with GAP in earlier work~\cite{Ugwumadu2026mapping} and
relaxed again here with ACE. The Stillinger--Weber gauge tests of Sec.~\ref{sec:gauge} use a small network
with original Stillinger--Weber parameters~\cite{Stillinger1985computer}: a 216-site diamond cell with 20 sites removed at random,
displaced randomly by 0.3~\AA\ and relaxed (196 atoms, 17\% three-fold). It is not meant as a model of a-Si. Stillinger--Weber is the one potential here whose site energies, and so every gauge, are known exactly, and a small defective cell suffices for that test; no result about a-Si rests on it. Coordination statistics of all models are in the Supplemental Material, Table~\textcolor{red}{S3}.

\paragraph*{Potentials.}
The atomic cluster expansion (ACE)~\cite{Drautz2019ace,Lysogorskiy2021pace} potential for Si--H of Rosset
and Deringer (SiH-ACE-25)~\cite{Rosset2026ace} is used throughout; GAP-18~\cite{Bartok2018silicon} (through
LAMMPS/QUIP) is used for si512 and si4096 (for si4096 from sparse finite differences; Supplemental Material, Sec.~\textcolor{red}{S4}). We also use the Keating valence force field~\cite{Keating1966effect} in the
form of Wooten, Winer and Weaire~\cite{Wooten1985computer},
$E=\sum_{\rm bonds}\frac{3\alpha}{16d^2}(r^2-d^2)^2+\sum_{\rm angles}\frac{3\beta}{8d^2}(\bm r\cdot\bm r'+d^2/3)^2$,
with $\alpha=2.965$~eV/\AA$^2$ (47.5~N/m), $\beta=0.285\alpha$ (13.5~N/m) and $d=2.35$~\AA, on the
bond topology of the input network. Each potential is used on its own relaxed structure. The
vibrational densities of states of all models are compared with inelastic neutron
scattering~\cite{Kamitakahara1987vibrational} in the Supplemental Material (Fig.~\textcolor{red}{S1}): the ACE and GAP-18 spectra
reproduce the measured one closely. The Keating spectrum does not: its transverse-optic-like (TO-like) peak lies near 48~meV,
about 10~meV below the measured band, and its transverse-acoustic-like (TA-like) band extends too high and fills the minimum near
30~meV (Kamitakahara \emph{et al.} also found the TA-like band of their Keating model too
high~\cite{Kamitakahara1987vibrational}).

\paragraph*{Dynamical matrices and conductivity.}
ACE dynamical matrices were computed by central finite differences (0.01~\AA) in LAMMPS~\cite{Thompson2022lammps},
on the GPU with the KOKKOS package. For BM10k we built a sparse dynamical matrix by finite
differences with graph colouring (atoms displaced together if more than $2r_c+1$~\AA\ apart), truncated
at 11~\AA\ with the acoustic sum rule restored; it reproduces the dense matrices of si512 and si4096 to
$10^{-8}$ and $3\times10^{-6}$. Unless stated otherwise we use $T=300$~K, quantum heat capacities and
$\eta=2$~K (0.17~meV); $\eta=10$~K (0.86~meV) is used as a second, smoother scale. For si4096 and
smaller models Eqs.~(\ref{eq:kappa})--(\ref{eq:terms}) are evaluated by dense diagonalization (about
15~min for $\kappa$ and $\zeta$, 17~min for the full gradient at 4096 atoms on one workstation). For
BM10k, $\kappa$ and $\zeta$ come from a double Chebyshev expansion of Eq.~(\ref{eq:zeta}) on the
GPU~\cite{Weisse2006kpm,Drabold1993maximum} (Supplemental Material, Sec.~\textcolor{red}{S6}), and $\dR$ from block-preconditioned conjugate
gradients on Eq.~(\ref{eq:dR}).

\paragraph*{Validation.}
For si512 the analytic gradient agrees with finite differences from an independent code to a median
relative error of $2\times10^{-5}$ over all sites (masses) and $10^{-6}$--$10^{-5}$ over 48 bonds. The
Euler rule, Eq.~(\ref{eq:euler}), holds to $10^{-6}$ in every $\eta$-broadened run, the three terms of
Eq.~(\ref{eq:decomp}) reconstruct $\km$ to $10^{-16}$, and $\sum_{\ell}\srot=0$ to $10^{-15}$. The
conjugate-gradient $\dR$ agrees with the exact pseudo-inverse of the same matrix to $10^{-6}$ (maximum
relative difference over all 4096 sites of si4096; mean $10^{-7}$), and truncation of the matrix at 11~\AA\
changes $\dR$ by at most 0.4\% (Pearson 0.99999 with the full dense result).

\paragraph*{Statistics.}
Most comparisons ask whether two maps order the sites alike, so we use the Spearman (rank) correlation
$S(x,y)$, which is insensitive to units, to monotonic rescaling and to the few extreme sites that are common
in $\km$. The partial correlation $S(x,y\,|\,\dR)$ correlates what remains of the ranks of $x$ and $y$ after
linear regression on the ranks of $\dR$; it is near zero if $\dR$ accounts for the relation. Neighbouring
sites are correlated, so we give 95\% intervals from a spatial block bootstrap (300 resamples of periodic cubic blocks: 64 for si4096 and BM10k, 27 for the 1000-atom models and 8 for
si512, so that blocks exceed the 8~\AA\ diameter of an averaging sphere), at single sites and for fields averaged over spheres of 4~\AA\ radius. As a yardstick for chance
agreement we shift one map rigidly through the periodic cell, which keeps its spatial autocorrelation, and recompute the
correlation (500 random translations, excluding those shorter than 8~\AA, twice the averaging radius; single-site values
are read at the nearest site, and 4~\AA\ fields are evaluated at the shifted points). Shuffling sites, the usual
alternative, destroys the autocorrelation and so underestimates chance correlations between smooth maps. The 95\% level of
the shifted correlation depends on the maps compared; for $\zeta$ it is 0.12 and 0.04 at single sites and 0.28 and 0.09 at
4~\AA\ (si512 and si4096; last row of Table~\ref{tab:robust}), for $\km$ against $\zeta$ in the 1000-atom models 0.15 at single sites and 0.29 at 4~\AA, and we quote it alongside other correlations where it matters; smaller correlations cannot be told from chance, and this is the sense in which we call one
``near zero''.

\section{Results}\label{sec:results}

\subsection{The heat-flow map is robust}\label{sec:robust}

For si4096 with ACE, $\kappa=1.218$~\WmK{} at $\eta=2$~K, of which diffusons contribute 0.956, propagons
0.241 and locons 0.021~\WmK{} (propagons below 10~meV, diffusons 10--56.5~meV, locons above 56.5~meV; a pair of modes
in different bands contributes half to each). The AF-gauge map of $\zeta$ and the map in the site-energy
gauge of the potential (obtained by probing the per-atom energies with finite displacements;
Supplemental Material, Sec.~\textcolor{red}{S9}) have Spearman correlation 0.92 at single sites and 0.984 over 4~\AA\ spheres. The
gauge uncertainty (the standard deviation over the two gauges divided by the spatial standard deviation of
their mean, after removing the uniform offset of result (i): $\kappa$ is 0.78\% lower in the site-energy gauge) is 23\% of the spatial spread of $\zeta$ at single sites and 9.5\% at 4~\AA\
(Fig.~\ref{fig:gauge}).

Table~\ref{tab:robust} compares maps across potentials and models. Two machine-learned potentials, ACE and GAP-18,
agree nearly as well as two gauges do (si512: $0.86$ at single sites and $0.92$ at 4~\AA\ for $\eta=2$~K; $0.92$
and $0.96$ at $\eta=10$~K; si4096: $0.89$ and $0.93$, and $0.90$ and $0.94$ at $\eta=10$~K).
GAP-18 gives a lower $\kappa$ than ACE at matched broadening: for si4096 1.009 against 1.218~\WmK{} at $\eta=2$~K and
0.867 against 0.948~\WmK{} at 10~K ($-17\%$ and $-9\%$; $-14\%$ and $-8\%$ for si512).
ACE and Keating, each on its own relaxed structure, agree at 0.84--0.89 for si4096. The independent BM10k
network reproduces the si4096 statistics: the relative spread of $\zeta$ (standard deviation over mean) is
0.12--0.13 in both, and $\zeta$ increases with the mean bond length of the atom (Spearman $+0.61$ to
$+0.63$ in BM10k, $+0.57$ in si4096). Thus, at the 4~\AA\ scale $\zeta$ is a robust structural descriptor of
where heat flows.

\begin{table}
\caption{Robustness of the heat-flow map $\zeta$ (Spearman correlations, single site and 4~\AA\
averages; $\eta=2$~K unless noted). Null: 95th percentile of the correlation with a cyclically shifted copy of the other map, the yardstick of Sec.~\ref{sec:models}.}
\label{tab:robust}
\begin{ruledtabular}
\begin{tabular}{lcc}
comparison & site & 4~\AA \\
\hline
gauge: AF vs site energies (ACE, si4096) & 0.92 & 0.984 \\
ACE vs GAP-18 (si512) & 0.86 & 0.92 \\
ACE vs GAP-18 (si512, $\eta=10$~K) & 0.92 & 0.96 \\
ACE vs GAP-18 (si4096) & 0.89 & 0.93 \\
ACE vs GAP-18 (si4096, $\eta=10$~K) & 0.90 & 0.94 \\
ACE vs Keating (si4096) & 0.84 & 0.89 \\
null (si512 / si4096) & 0.12 / 0.04 & 0.28 / 0.09 \\
\end{tabular}
\end{ruledtabular}
\end{table}

\begin{table*}[!t]
\caption{Anti-correlation of the mass map with the heat-flow map and its mediation by local softness.
$S(x,y)$: Spearman rank correlation of maps $x$ and $y$, for fields averaged over 4~\AA\ spheres, with 95\% block-bootstrap intervals.
Last column: $S(\km,\zeta\,|\,\dR)$, the partial correlation of $\km$ and $\zeta$ at fixed $\dR$, that is, the correlation of what remains of the two maps once the part of each that follows $\dR$ is removed (Sec.~\ref{sec:models}). $T=300$~K, quantum heat
capacities; $\dR$ classical. $^a$Single sites (455 sampled sites of the relaxed model; $\km$ not computed).}
\label{tab:anti}
\begin{ruledtabular}
\begin{tabular}{lccccc}
model, potential, $\eta$ & $\kappa$ (\WmK{}) & $S(\km,\zeta)$ & $S(\zeta,\dR)$ & $S(\km,\dR)$ & $S(\km,\zeta\,|\,\dR)$ \\
\hline
si4096, ACE, 2~K & 1.218 & $-0.41$ $[-0.49,-0.34]$ & $-0.71$ $[-0.73,-0.68]$ & $+0.61$ $[+0.53,+0.66]$ & $+0.02$ $[-0.05,+0.09]$ \\
si4096, Keating, 2~K & 1.734 & $-0.66$ $[-0.70,-0.62]$ & $-0.86$ $[-0.88,-0.84]$ & $+0.76$ $[+0.72,+0.80]$ & $+0.00$ $[-0.10,+0.09]$ \\
si4096, GAP-18, 2~K & 1.009 & $-0.31$ $[-0.37,-0.24]$ & $-0.66$ $[-0.70,-0.61]$ & $+0.47$ $[+0.39,+0.54]$ & $+0.00$ $[-0.06,+0.08]$ \\
R1, ACE, 2~K & 8.15 & $-0.51$ $[-0.67,-0.35]$ & $-0.88$ $[-0.93,-0.80]$ & $+0.50$ $[+0.32,+0.66]$ & $-0.15$ $[-0.30,+0.06]$ \\
R2, ACE, 2~K & 1.72 & $-0.55$ $[-0.66,-0.42]$ & $-0.73$ $[-0.82,-0.59]$ & $+0.58$ $[+0.44,+0.69]$ & $-0.24$ $[-0.37,-0.09]$ \\
R3, ACE, 2~K & 1.74 & $-0.43$ $[-0.61,-0.22]$ & $-0.79$ $[-0.86,-0.70]$ & $+0.54$ $[+0.32,+0.68]$ & $-0.01$ $[-0.17,+0.17]$ \\
si512, ACE, 2~K & 1.091 & $-0.25$ $[-0.38,-0.10]$ & $-0.77$ $[-0.83,-0.71]$ & $+0.32$ $[+0.22,+0.41]$ & $-0.02$ $[-0.20,+0.20]$ \\
si512, ACE, 10~K & 0.972 & $-0.32$ $[-0.44,-0.16]$ & $-0.81$ $[-0.86,-0.74]$ & $+0.51$ $[+0.38,+0.61]$ & $+0.20$ $[+0.03,+0.36]$ \\
si512, GAP-18, 2~K & 0.944 & $-0.13$ $[-0.27,-0.00]$ & $-0.65$ $[-0.76,-0.52]$ & $+0.05$ $[-0.08,+0.18]$ & $-0.13$ $[-0.26,+0.00]$ \\
si512, GAP-18, 10~K & 0.892 & $-0.00$ $[-0.13,+0.14]$ & $-0.67$ $[-0.78,-0.52]$ & $+0.31$ $[+0.21,+0.43]$ & $+0.29$ $[+0.20,+0.38]$ \\
BM10k, ACE, 2~K$^a$ & 1.231 & --- & $-0.63$ $[-0.68,-0.55]$ & --- & --- \\
\end{tabular}
\end{ruledtabular}
\end{table*}

\subsection{The sensitivity map is anti-correlated with the heat-flow map}\label{sec:anti}

\begin{figure*}[!t]
\includegraphics[width=\textwidth]{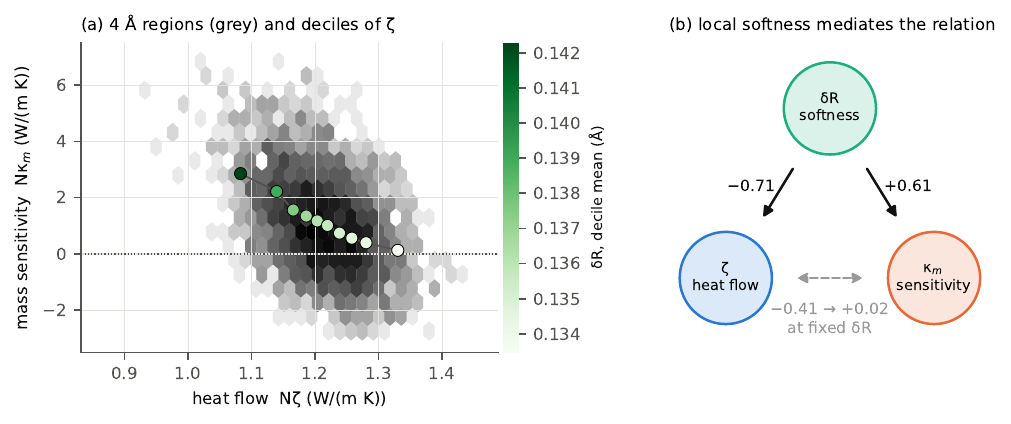}
\caption{Mass sensitivity against heat flow for si4096 (ACE, $\eta=2$~K, fields averaged over 4~\AA\ spheres and
multiplied by $N$). (a) Density of the regions (grey, logarithmic) and the mean $\km$ in each tenth of the
regions ordered by $\zeta$, coloured by their mean vibrational amplitude $\dR$. (b) Local softness mediates
the relation: Spearman correlations between the three maps, and the partial correlation of $\km$ and
$\zeta$ at fixed $\dR$.}
\label{fig:scatter}
\end{figure*}

Figure~\ref{fig:scatter}(a) compares the mass sensitivity $\km$ with $\zeta$ for si4096. The two are
anti-correlated: Spearman $-0.31$ $[-0.36,-0.26]$ at single sites and $-0.41$ $[-0.49,-0.34]$ at 4~\AA\
(95\% intervals). Table~\ref{tab:anti} collects the models at $\eta=2$~K; $\eta=10$~K is in the Supplemental Material (Table~\textcolor{red}{S4}). The anti-correlation appears with all three potentials and in every network of 1000 atoms or more: it is stronger with Keating ($-0.66$ at 4~\AA) and in the
paracrystalline models ($-0.43$ to $-0.55$), holds for classical as well as quantum heat capacities, and
for broadening up to $\eta=10$~K. It is not driven by outliers: removing the top 1\% of sites leaves it
unchanged. With GAP-18 it is weaker than with ACE but well clear of chance at 4096 atoms ($-0.31$ $[-0.37,-0.24]$
at 4~\AA\ and $\eta=2$~K against a null of 0.09; $-0.28$ at single sites). At 512 atoms it is weak with both
potentials: for GAP-18 $-0.13$ at $\eta=2$~K and $-0.00$ at 10~K at 4~\AA, below the shifted nulls of 0.30 and 0.23, and for ACE
$-0.25$ (null 0.21) against $-0.41$ at 4096 atoms. The weakness at 512 atoms is thus common to both potentials, as a
finite-size effect would be, and is not a property of GAP-18. In
the 32\,768-atom replica of si4096 it is stronger still ($-0.56$ at 300~K; Sec.~\ref{sec:temperature}).
Thus the anti-correlation strengthens as the cell supports more low-frequency modes: from si512 to
si4096, and in the 32\,768-atom replica of si4096, which adds longer-wavelength modes to the same network.
We return
to this in Sec.~\ref{sec:discussion}.

Since $\km$ has both signs (38\% of the sites of si4096 are negative at $\eta=2$~K), the anti-correlation is
first of all a statement about \emph{sign}. In every model the 10\% of regions with the least heat flow have positive mean
$\km$: a heavier atom there lowers $\kappa$. The 10\% with the most heat flow have negative or small mean
$\km$. The magnitude follows less closely. Over 4~\AA\ the regions with the least heat flow have the larger
mean $|\km|$ (2.3 times that of the most heat-carrying regions for si4096 with ACE, 2.4 with Keating, 1.3--6
for R1--R3), and Spearman$(|\km|,\zeta)$ is $-0.18$ (ACE) and $-0.41$ (Keating); but at single sites the
magnitude is nearly uncorrelated with $\zeta$ ($-0.02$ for ACE at $\eta=2$~K), and for R2 and for GAP-18 at 512 atoms it
is uncorrelated at 4~\AA\ as well ($-0.14$ for GAP-18 at 4096 atoms) (Supplemental Material, Table~\textcolor{red}{S5}). Thus ``where the response is
large'' and ``where heat flows little'' coincide only on average; what the heat flow sets reliably is the
\emph{sign} of the response. The picture is plain in three dimensions, where the two maps occupy largely
different parts of the network: the 12\% of
the volume where the smoothed $\zeta$ is largest and the 12\% where $\km$ is largest share 0.27\% of the
volume, five times less than independent maps would [Fig.~\ref{fig:maps}(a)].

The extreme values of $\km$ at $\eta=2$~K belong to under-coordinated atoms with quasi-localized
resonances: 4 of the 11 most sensitive sites of si4096 are among its 5 three-fold atoms, and their
sensitivity is dominated by a single mode (for example at 7.3~meV with participation ratio 0.06).
These resonances decorate the map but do not cause the anti-correlation: the Keating network keeps the
four-bonded topology of its input and has no coordination defects, yet its anti-correlation is stronger; a defect-free
ACE model~\cite{Rosset2026defectfree} would test this further.

\begin{figure*}[!t]
\includegraphics[width=\textwidth]{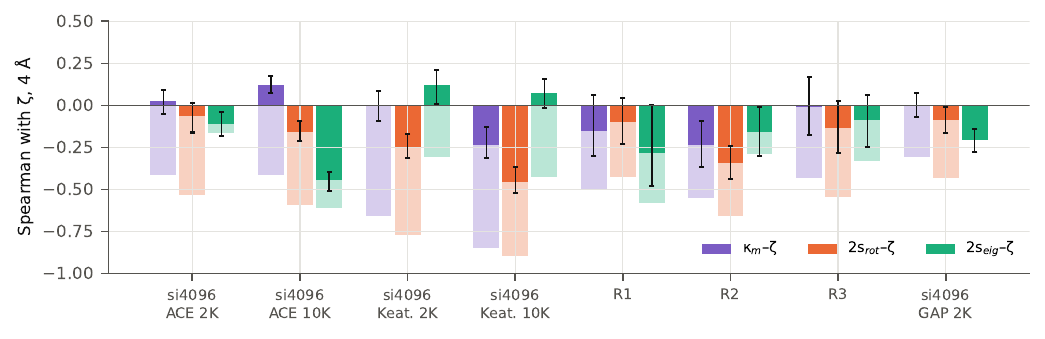}
\caption{Which term is mediated: Spearman correlation (4~\AA) of each term of Eq.~(\ref{eq:decomp}) with $\zeta$,
without conditioning (pale) and at fixed $\dR$ (solid), for si4096 (ACE and Keating, $\eta=2$ and 10~K), R1--R3 (ACE, 2~K) and
si4096 (GAP-18, 2~K); bars are 95\% block-bootstrap
intervals. The other models of Table~\ref{tab:anti} are not shown. The correlations of the three maps themselves are in Table~\ref{tab:anti} and
Fig.~\ref{fig:scatter}(b).}
\label{fig:mediation}
\end{figure*}

\subsection{Mechanism: local softness and mode mixing}\label{sec:mechanism}

\paragraph*{Local softness mediates the anti-correlation.}
The heat-flow map is strongly anti-correlated with the local vibrational amplitude $\dR$
[Table~\ref{tab:anti}, Fig.~\ref{fig:scatter}(b), and Supplemental Material, Table~\textcolor{red}{S4}]: $-0.61$ to $-0.90$ at 4~\AA\ in every model for which $\km$ was computed, and
$-0.63$ in BM10k (single sites). The sensitivity map is positively
correlated with $\dR$: $+0.31$ to $+0.90$, including $+0.47$ for GAP-18 at 4096 atoms ($+0.05$ for GAP-18 at 512 atoms and $\eta=2$~K). Soft, large-amplitude regions carry little heat, and there $\km$ is
large and positive: a heavier atom lowers $\kappa$ most. At $\eta=2$~K and fixed $\dR$ the partial correlation of $\km$ and
$\zeta$ falls to $+0.02$, $+0.00$ and $+0.00$ for si4096 with ACE, Keating and GAP-18, and lies between $-0.24$ and $+0.20$ in
every ACE and Keating case and for GAP-18 at 4096 atoms, against $-0.22$ to $-0.85$ without conditioning (Table~\ref{tab:anti} and
Supplemental Material, Table~\textcolor{red}{S4}). At $\eta=10$~K the partial correlation becomes positive for both
machine-learned potentials ($+0.12$ $[+0.06,+0.18]$ for ACE, $+0.18$ $[+0.12,+0.25]$ for GAP-18; Supplemental
Material, Table~\textcolor{red}{S4}), so at that broadening the amplitude more than accounts for the anti-correlation. The local low-frequency weight
$w_{\rm low}$ is an equally good mediator; conditioning on both adds nothing over either alone,
so both measure the same local property. In spectral terms, the 10\% of sites with the largest $\dR$
and the 10\% with the smallest $\zeta$ have about twice the weight below 12.4~meV of their opposite
numbers (Supplemental Material, Sec.~\textcolor{red}{S5}).

The independent BM10k network confirms the link without any eigenvectors. With $\dR$ from sparse
linear solves and $\zeta$ from the Chebyshev expansion at sampled sites (455, 522 and 547 in the three
states), the single-site $S(\zeta,\dR)=-0.63$ $[-0.68,-0.55]$ for the relaxed model and $-0.68$ and $-0.71$
after the 800 and 1000~K anneals. Restricted
to four-fold sites it is $-0.47$ to $-0.49$, so the relation is not carried by defects (Supplemental Material, Sec.~\textcolor{red}{S6}).

\paragraph*{Which term is mediated.}
Equation~(\ref{eq:decomp}) shows how (Fig.~\ref{fig:mediation}). Of the covariance between $\km$ and $\zeta$ at 4~\AA, the current
term $4\zeta$, positively correlated with $\zeta$ by construction, opposes the total (share $-9$ to $-108\%$ in
si4096 and R1--R3 at $\eta=2$ and 10~K), the eigenvalue-shift term carries 7--38\%, and the rotation term
80--171\%, overcompensating the current term in each of these models; with GAP-18 at 4096 atoms the pattern is the same but the shares are larger ($-68$/$-225$\%, $+20$/$+44$\%, $+148$/$+281$\% at $\eta=2$/10~K), because the total covariance is smaller; in the 512-atom cells, where the total
covariance is small, all three shares are larger (Supplemental Material, Table~\textcolor{red}{S6}). Conditioning on $\dR$
removes most but not all of the rotation term's anti-correlation with $\zeta$: for si4096 with ACE at
$\eta=2$~K it falls from $-0.53$ to $-0.06$, and for R1--R3 from $-0.42$/$-0.66$/$-0.55$ to
$-0.10$/$-0.34$/$-0.13$. A residual remains with Keating ($-0.77\to-0.25$ at 2~K, $-0.90\to-0.46$ at
10~K). Local softness thus sets the magnitude of the effect, and eigenvector rotation is the microscopic
route; in the stiffest, most uniform network (Keating) part of the rotation term carries information that
$\dR$ does not.

\begin{figure*}[!t]
\includegraphics[width=.9\textwidth]{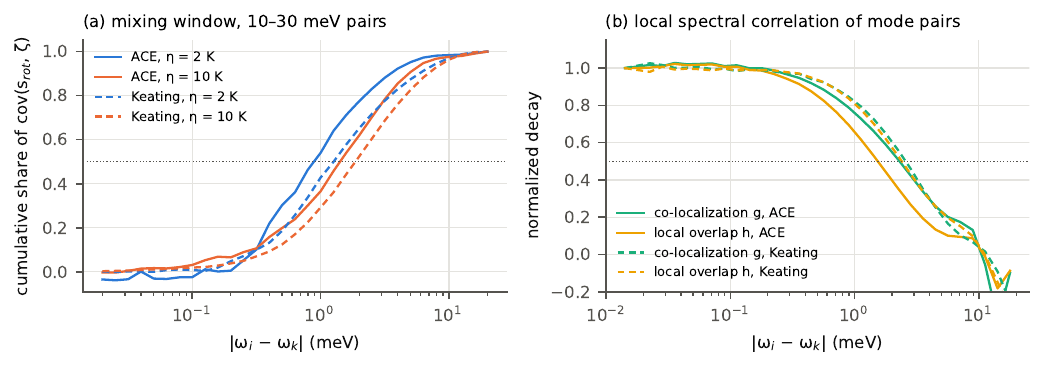}
\caption{Origin of the mixing window, si4096. (a) Cumulative share of the covariance between $\srot$ and
$\zeta$ (4~\AA\ averages) carried by pairs of 10--30~meV modes closer than $|\omega_i-\omega_k|$.
(b) Co-localization $g$ and local overlap strength $h$ of the same mode pairs, normalized to 1 at small
separation and 0 at 5--20~meV.}
\label{fig:window}
\end{figure*}

\paragraph*{The mixing window.}
Resolving $\srot$ into mode pairs shows that the anti-correlation is carried by pairs of modes about
1--2~meV apart, mostly in the 10--30~meV band of low diffusons (54--74\% of the covariance from pairs whose mean frequency
lies in that band); modes above
45~meV contribute nothing. The window is set mainly by the network, not by the broadening: at $\eta=2$~K (0.17~meV) pairs within
$3\eta$ carry only 26--29\% of the covariance. It is set by the local spectral correlation of the modes
[Fig.~\ref{fig:window}]. For the
4286 (ACE) and 3719 (Keating) modes of the 10--30~meV band we computed, as functions of
$|\omega_i-\omega_k|$, the co-localization $g_{ik}=N\sum_{\ell}p_i({\ell})p_k({\ell})-1$ and the local overlap strength
$h_{ik}=N\sum_{\ell}c_{ik}({\ell})^2$, which measure whether two modes live on the same atoms. Both decay to half
their value over 1.5--2.6~meV ($h$: 1.5~meV for ACE, 2.4~meV for Keating; $g$: 2.3 and 2.6~meV), independent
of $\eta$. The rotation covariance of these pairs reaches half its total at 0.9 and 1.4~meV for ACE ($\eta=2$ and 10~K)
and at 1.3 and 1.9~meV for Keating: a fivefold change of $\eta$ widens the window by only half, and the
potential with the broader local overlap has the broader window. By contrast, the correlations of the
current matrix elements ($\langle|P_{ik}|^2\rangle$, the $F$-weighted overlap of $P$ rows, and $N_{ik}^2$)
show no common decay scale (Supplemental Material, Table~\textcolor{red}{S7}). The mixing window is therefore the energy range over
which two vibrational modes share the same atoms: a local mass change can only mix modes that overlap
locally, and in a-Si that requires them to be within 1--2~meV. This is, in our view, the vibrational form of
the island picture of localized electronic states in disordered networks, in which eigenstates close in
energy are built from the same spatial islands~\cite{Dong1998bandtail,Ludlam2005universal}.

The scale is suggestive of a Thouless energy~\cite{Chalker1988scaling,Mirlin2000statistics}. The AF
diffusivity averaged over the 10--30~meV band is 0.94~mm$^2$/s for ACE and 1.60~mm$^2$/s for Keating, a
ratio of 1.7, while the local-overlap widths differ by a factor of 1.6 (1.5 and 2.4~meV). Writing the width
as $E_w=\hbar D/L^2$ gives the same length for both potentials: $L=6.4$ and 6.6~\AA\ from the overlap
width, and 8.3 and 9.1~\AA\ from the half-point of the rotation covariance at $\eta=2$~K. These lengths are
2.5--4 bond lengths, close to the Ioffe--Regel scale, so the diffusive form $\hbar D/L^2$ is an
order-of-magnitude statement. Modes closer in
energy than $\hbar D/L^2$ remain correlated over a region about 0.6--0.9~nm across, and only such pairs are
mixed by a local perturbation. It is reasonable to
conjecture that the window is a Thouless energy of this kind; the two potentials are consistent with it,
but they do not prove it.

\begin{figure}[!t]
\includegraphics[width=\columnwidth]{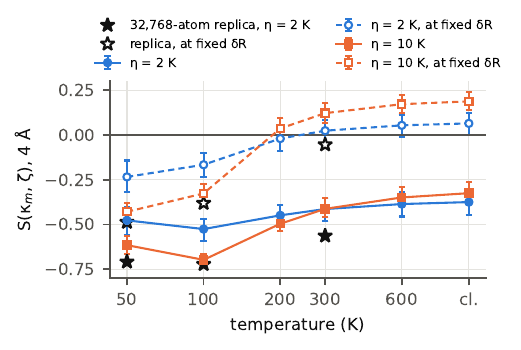}
\caption{Spearman correlation of $\km$ and $\zeta$ at 4~\AA\ for si4096 (ACE) against temperature
(``cl.'': classical heat capacities), without conditioning (filled) and at fixed classical $\dR$ (open),
for $\eta=2$ and 10~K, with 95\% block-bootstrap intervals. Stars: the 32\,768-atom replica of si4096
($\eta=2$~K), without conditioning (filled) and at fixed $\dR$ (open).}
\label{fig:temperature}
\end{figure}

\begin{figure*}[!t]
\includegraphics[width=.8\textwidth]{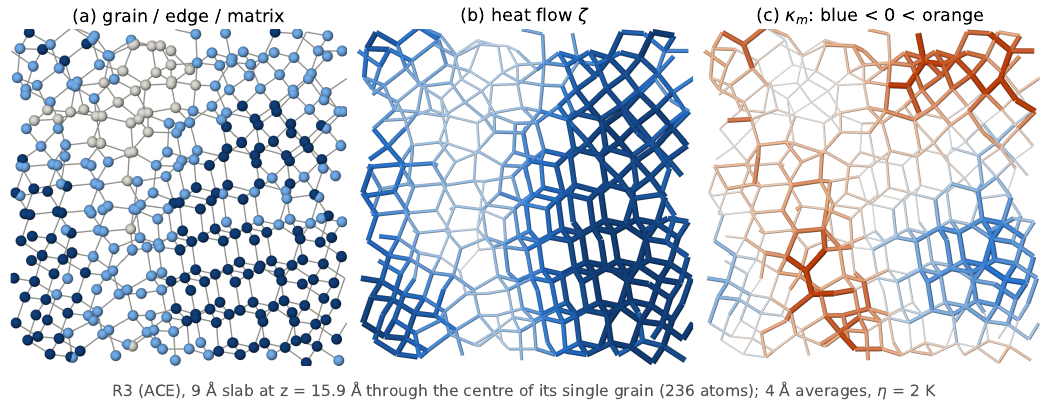}
\caption{The paracrystalline model R3 (ACE): a 9~\AA\ slab through the centre of its single grain (236 atoms,
24\% of the cell). (a) Atoms classified by diamond-structure identification (OVITO): crystalline (cubic or
hexagonal diamond, dark), grain edge (first and second neighbours of crystalline atoms, light) and matrix
(grey). (b) Heat-flow map $\zeta$ and (c) mass sensitivity $\km$ on a diverging scale centred at zero (blue
negative, orange positive), both averaged over 4~\AA\ spheres, $\eta=2$~K. The grain carries more heat and has
negative $\km$; the matrix carries less and has positive $\km$.}
\label{fig:para}
\end{figure*} 

\subsection{Temperature dependence}\label{sec:temperature}

All points from 50~K to the classical limit, at $\eta=2$ and 10~K, were computed from a single
diagonalization of si4096 (ACE); the Euler rule closes to $10^{-6}$ at each. The anti-correlation persists
at all temperatures (Fig.~\ref{fig:temperature}) and is strongest at low $T$: Spearman at 4~\AA\ is
$-0.53$ at 100~K and $-0.38$ in the classical limit ($\eta=2$~K), and $-0.70$ and $-0.33$ at
$\eta=10$~K. The rotation term carries 93--122\% of the covariance at every temperature, and half of it
comes from pairs less than 1~meV apart at $\eta=2$~K (1.4--1.7~meV at $\eta=10$~K) at all $T$. The mixing
window is thus a property of the spectrum and the broadening, not of the thermal population. The map of
$\zeta$ is stable in $T$ (Spearman 0.97 between 100 and 600~K at $\eta=10$~K); the map of $\km$ less so
(0.79), because at low $T$ the soft modes weigh more.

Above 200~K the classical $\dR$ largely mediates the anti-correlation (partial correlations $-0.02$ to
$+0.19$). At 50--100~K it mediates only part: the partial correlation is $-0.17$ to $-0.43$. Zero-point
motion does not account for the remainder. The quantum amplitude,
$\dR_{\ell}^2=\sum_ip_i({\ell})(\hbar/2m\omega_i)\coth(\hbar\omega_i/2k_BT)$, ranks sites almost as the classical
one does (Spearman 0.97 at 50~K, 0.996 at 100~K), and conditioning on it leaves partial correlations of
$-0.21$ to $-0.50$. Nor is the residual carried by the heat-capacity channel of the eigenvalue term (heavier atoms lower
frequencies and raise the populations of modes with $\hbar\omega$ of a few $k_BT$): this channel is 31\% of
$\sum_{\ell}\seig$ at 50~K, but it is positively correlated with $\zeta$, and removing it from $\km$ leaves
the residual unchanged ($-0.23$). At 100~K, conditioning jointly on $\dR$ and on the local weight in the
thermally active band ($k_BT\le\hbar\omega\le5k_BT$) reduces the residual to $-0.08$; at 50~K it does
not ($-0.23$).

The residual is \emph{not} removed by adding longer-wavelength modes. We repeated the analysis on the 32\,768-atom replica (Supplemental Material, Sec.~\textcolor{red}{S13} and Table~\textcolor{red}{S8})
of si4096, computed as eight independent $k$-blocks (the current operator does not couple different $k$),
whose lowest modes reach down to 2.0~meV instead of 3.9~meV. At 300~K the replica confirms the main
result: the anti-correlation is stronger ($-0.56$ against $-0.41$), and at fixed $\dR$ it again largely
disappears ($-0.05$ $[-0.12,0.00]$). At low temperature the residual grows when the longer-wavelength modes are included: at fixed $\dR$ it
is $-0.38$ at 100~K (cell $-0.17$) and $-0.49$ $[-0.55,-0.41]$ at 50~K (cell $-0.23$). Thus the additional
long-wavelength modes, which carry much of the heat when the thermal energy $k_BT$ is only a few meV, contribute an
anti-correlation that local softness does not capture. The picture is easily stated: in the diffuson
regime the response map is local and is set by the soft regions; when propagons carry the heat it
acquires a component that local softness does not capture. Three simple explanations fail. Restricting the AF weight to pairs of
propagons (both below 10~meV) splits $\km$ exactly; at 50~K these pairs carry 54\% of $\kappa$ and 78\%
of the covariance of $\km$ with $\zeta$, but the part not explained by $\dR$ is about the same in the
propagon and the remaining contributions ($-0.40$ and $-0.36$). Nor is it softness on a longer scale:
averaging $\dR$ over spheres of 6, 8 and 12~\AA\ makes the residual larger ($-0.56$, $-0.63$, $-0.68$),
so softness on a longer scale is not the missing mediator; conditioning jointly on $\dR$ at 4 and 12~\AA\ leaves
$-0.49$ at 50~K, so the residual is not carried by softness on either scale. Nor is it an inconsistency between quantum
heat capacities in the maps and a classical amplitude: with the quantum amplitude, including zero-point
motion, the replica residual is larger still ($-0.60$ at 50~K and $-0.44$ at 100~K), while at 300~K the two
amplitudes give the same result. Conditioning on both amplitudes together, which are different moments of the same
local mode weights, roughly halves the residual ($-0.29$ at 50~K, $-0.19$ at 100~K), far more than a second field with
the same spatial structure but displaced at random does (Supplemental Material, Sec.~\textcolor{red}{S13}); this suggests that
the shape of the local vibrational spectrum, not a single moment of it, mediates the correlation. The component lies in the rotation term: at fixed $\dR$, $2\srot$ keeps a
partial correlation of $-0.49$ with $\zeta$ at 50~K but only $-0.10$ at 300~K, while the eigenvalue term is
unchanged ($-0.28$ and $-0.29$). Thus at low temperature the mixing of modes is no longer set by local softness alone: nearly half of the
residual follows the shape of the local vibrational spectrum, and the rest remains open.

\subsection{Paracrystalline silicon}\label{sec:paracrystal}

Paracrystalline silicon, amorphous silicon containing crystalline grains of nanometre
size~\cite{Voyles2001paracrystalline,Nakhmanson2001realistic}, has been inferred from fluctuation electron
microscopy~\cite{Treacy2012local}, and the models R1--R3 of Rosset \emph{et al.}~\cite{Rosset2025paracrystallinity}
make it concrete. Identifying cubic and hexagonal diamond environments (with OVITO), we find crystalline
fractions of 18\% (R2), 24\% (R3) and 93\% (R1, essentially polycrystalline), against 0.1\% in the DTW and
Barkema--Mousseau networks. The two paracrystalline models conduct about 40\% better than a-Si (1.72 and
1.74~\WmK{} for R2 and R3 against 1.22~\WmK{}, $\eta=2$~K), but with two models the trend with crystalline
fraction cannot be resolved; R1 (8.15~\WmK{}) is limited by the broadening and is not comparable. The maps show where the
difference arises [Fig.~\ref{fig:para}; profiles and regional averages in the Supplemental Material, Fig.~\textcolor{red}{S2} and Table~\textcolor{red}{S9}].

Moving from the amorphous matrix into a grain, the local heat flow rises steadily (for R3, $N\zeta$ from 1.60 in the matrix to 2.06~\WmK{} in grain interiors), the
vibrational amplitude falls (0.1286 to 0.1225~\AA), and the mass sensitivity changes sign (from $+3.1$ to
$-3.8$~\WmK{}). Thus the grains conduct about 30\% more heat per atom than the matrix, but, at 24\% of the
atoms, they do not carry most of it; the grain boundaries and the amorphous matrix control the response.
The nearly polycrystalline R1 is different: its grain interiors keep a positive $\km$ ($+2.1$, against $+108$ in its
thin matrix), so the change of sign belongs to grains embedded in a matrix, not to crystallinity alone. A plausible
reason is that the grains of R1 are large enough to carry propagon-like modes, which respond to a mass change
differently from diffusons.
The negative sensitivity of grain interiors is a first-order statement only. Exact recalculation shows
that $\kappa$ is nearly stationary with respect to grain-interior masses, as for the crystal at the
interface (Sec.~\ref{sec:interface}): 5\% heavier raises $\kappa$ of R3 by 0.3\% (linear 0.5\%), but 20\%
heavier lowers it by 0.5\% (linear $+1.9\%$). The same relation of softness to heat flow
and to sensitivity that holds within a-Si holds here between the two phases, which gives a thermal
signature of paracrystallinity: it is the disordered fraction, not the crystalline one, that sets how
$\kappa$ responds to change.

\subsection{Material design: additivity and substituent placement}\label{sec:design}

The sensitivity map is a first-order response. We tested how far it predicts finite changes by exact
recalculation of $\kappa$ for si4096 after each perturbation, at $\eta=2$ and 10~K; where sites are
chosen from the map, the $\eta=10$~K map is used. Details and the full results are in the Supplemental Material, Sec.~\textcolor{red}{S15} and Table~\textcolor{red}{S10}.

\paragraph*{Isotopes and bonds.} Isotope substitution ($^{30}$Si) and bond-stiffness changes of
$\pm5\%$ are nearly linear over regions of up to 46 atoms (6~\AA\ radius): the ratio of the exact change to the
linear prediction is 1.10--1.21 for isotopes and 0.93--1.06 for bonds. Bond changes of $\pm20\%$ are
not (0.67 and 1.17). Pairs of substituents act additively to within 2\% beyond 5~\AA; nearest
neighbours interact at the 7--9\% level.

\begin{figure}
\includegraphics[width=\columnwidth]{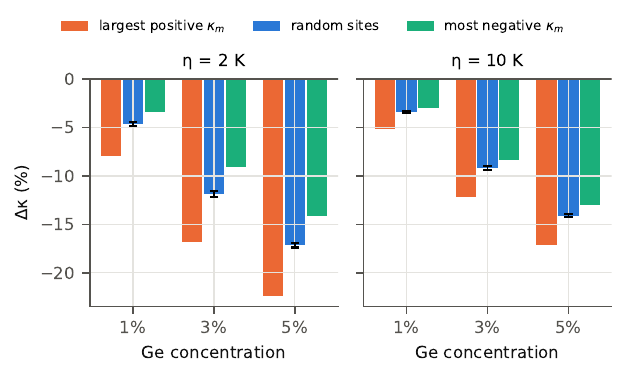}
\caption{Exact change of $\kappa$ when Ge replaces 1, 3 or 5\% of the Si atoms of si4096 (ACE), at sites at
least 5~\AA\ apart chosen as the sites of largest positive $\km$, at random (mean and standard
deviation of five sets), or as the sites of most negative $\km$ ($\eta=10$~K map; mass-only Ge).}
\label{fig:design}
\end{figure}

\paragraph*{Heavy substituents.} We model germanium by its mass alone ($m=72.6$, Si force constants, no
relaxation): chemical bonding, structural relaxation and species-dependent force constants are excluded, so
the results are not predictions for real Si--Ge alloys. This mass-only Ge is far outside the linear regime: in compact
regions the exact change is 2.3--4.0 times the prediction at $\eta=2$~K, and every Ge substitution lowers
$\kappa$, including where the map predicts a rise. In compact clusters the effect does not depend on
where the cluster sits: at $\eta=10$~K a 46-atom Ge region lowers $\kappa$ by 2.33, 2.38 and 2.42\% when
centred on the sites of largest positive $\km$, random sites, and the sites of most negative $\km$. A compact cluster scatters as a single
heavy defect.
For dispersed substituents the map is a useful guide (Fig.~\ref{fig:design}). We placed Ge on 1, 3 and
5\% of the sites of si4096, selected by the signed rank of $\km$ in the $\eta=10$~K map, subject to a minimum
pair separation of 5~\AA: the sites with the largest positive $\km$ (where a heavier atom is predicted to
lower $\kappa$ most), those with the most negative $\km$, or random sites (five sets). At 5\% and $\eta=2$~K,
$\kappa$ falls by 22.4\% for the most positive sites, $17.2\pm0.2\%$ for random sites and 14.1\% for
the most negative sites: guided placement lies far outside the spread of the random sets (23 times their
standard deviation, from five sets), and the two guided placements span 8 percentage points of $\kappa$. The ordering holds at every
concentration and at $\eta=10$~K (17.2, $14.1\pm0.2$ and 13.0\% at 5\%). The separation constraint
matters: random sets without it, where Ge atoms can cluster, lower $\kappa$ by 18.7\% at 5\%. Thus, for
dispersed heavy substituents the map is a useful ranking tool, while for compact clusters only the number
of substituted atoms matters.

\subsection{An amorphous/crystalline interface}\label{sec:interface}

\begin{figure*}
\includegraphics[width=\textwidth]{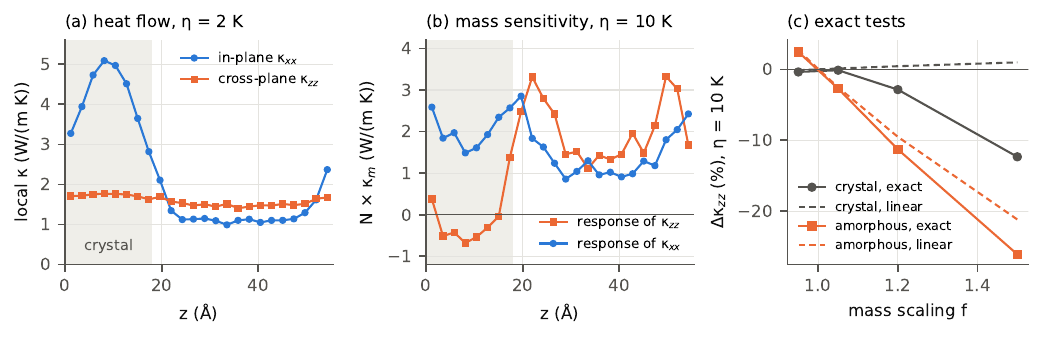}
\caption{The a-Si/c-Si interface model (2022 atoms, ACE). (a) Local in-plane and cross-plane
conductivity ($N$ times the mean $\zeta$ in 2.3~\AA\ slices; $\eta=2$~K); shaded: crystal. (b) Mass
sensitivity of the cross-plane and in-plane conductivity ($\eta=10$~K). (c) Exact change of $\kappa_{zz}$
when all crystal or all amorphous masses are scaled by $f$ (solid), against the linear prediction from the
map (dashed).}
\label{fig:interface}
\end{figure*}

As an application we consider heat flow across an amorphous/crystalline interface: a 2022-atom periodic
cell with a crystalline layer ($0\le z<18$~\AA, 650 atoms) between amorphous regions (1372 atoms), so
that the cell contains two interfaces. The model is that of Ref.~\cite{Bernstein1998interface}, relaxed with
GAP in Ref.~\cite{Ugwumadu2026mapping}; relaxing it again
with ACE moves atoms by 0.12~\AA\ rms and leaves the crystal essentially unchanged (0.04~\AA\ rms,
0.11~\AA\ at most). The crystal is defect free; 3.6\% of the amorphous atoms are three- or five-fold, more
than in the bulk models, so this is a model interface rather than an annealed one. The gauge uncertainty of $\zeta$ is 1\% of its spread at 4~\AA\ (4\% at single sites), small here because the spread includes the contrast between crystal and amorphous regions; within the amorphous region more than 4~\AA\ from the interfaces it is 7\% at 4~\AA, close to the bulk value.

\paragraph*{Where heat flows.} The in-plane conductivity is $\kappa_{xx}=2.25$~\WmK{} and the cross-plane
conductivity $\kappa_{zz}=1.59$~\WmK{} ($\eta=2$~K); directional conductivities $\kappa_{\alpha\alpha}$ are three times the
$\alpha$ term of Eq.~(\ref{eq:kappa}). The in-plane local conductivity is about 5~\WmK{} in the
crystal (limited by the cell and by $\eta$) and 1.0--1.2~\WmK{} in the amorphous region, the bulk a-Si value,
with transitions about 5~\AA\ wide [Fig.~\ref{fig:interface}(a)]. The cross-plane profile of $\zeta_{zz}$ is
nearly flat (1.4--1.8~\WmK{}). This is just what one expects for layers conducting in series. In a steady state the
same cross-plane flux passes through every plane, so the share of $\kappa_{zz}$ carried by a slab of
thickness $d$ is close to $(d/L)\kappa_{zz}$ wherever the resistance lies, up to terms from bonds that cross
the faces of the slab. As for resistors in series, which carry the same current, the current map \emph{cannot}
locate the region that limits cross-plane transport. Modal analysis of interface conductance~\cite{Gordiz2015formalism}
decomposes the conductance over pairs of modes; the question here is different: where a change would alter it.

\paragraph*{Where to change it.} The sensitivity map of $\kappa_{zz}$ can: it identifies the region where
local perturbations most affect cross-plane conductivity. Its mass sensitivity peaks about 4--6~\AA\
into the amorphous side of both interfaces, is moderate in the amorphous interior, and is small in the crystal [Fig.~\ref{fig:interface}(b)]. Of the 40 bonds whose stiffening most raises
$\kappa_{zz}$, none lies in the crystal and 25 lie within 8~\AA\ of an interface on the amorphous side (against 11
expected for bonds chosen at random). Low diffusons straddle the interfaces, low-frequency propagons lean
toward the amorphous side, and there is no band of interface-localized modes; within each region the
rattler statistics of the bulk reappear (Supplemental Material, Sec.~\textcolor{red}{S16}).

\paragraph*{Exact tests.} We scaled all crystal masses, or all amorphous masses, uniformly and recomputed
$\kappa_{zz}$ exactly (Table~\ref{tab:interface}). The amorphous side behaves as the map predicts:
lighter amorphous atoms raise $\kappa_{zz}$ (by 2.5\% for a 5\% mass reduction, against 2.7\% predicted),
and the linear prediction holds to within 20--25\% even for a 50\% mass increase ($-26\%$ against $-21\%$).
The crystal response is dominated by second-order terms: $\kappa_{zz}$ is stationary near the natural
crystal mass, so changes of either sign lower it, and a heavier crystal does not improve acoustic
matching ($-12\%$ for 50\% heavier). A plausible reason is that the crystal layer has near-degenerate modes: the first-order map is exact,
but its range of validity is set by their splittings, which a uniform mass change of a few per cent already
exceeds. The crystal is also where a uniform broadening is least physical, since its local conductivity is
limited by the cell and by $\eta$, so the curvature may be in part an artifact of the broadening. The cross-plane conductivity is
controlled from the amorphous side, and the response map locates that region. Two caveats apply: this is
a model interface with a defective amorphous region, and the crystal is treated at the $\Gamma$ point of a
periodic cell.

\begin{table}
\caption{Exact change of $\kappa_{zz}$ and $\kappa_{xx}$ ($\eta=10$~K) of the interface model when all
crystal or all amorphous masses are scaled by $f$, with the linear prediction
$-\sum_{{\ell}\in\text{region}}s_m({\ell})\ln f$ from the direction-resolved mass map.}
\label{tab:interface}
\begin{ruledtabular}
\begin{tabular}{lcccc}
region, $f$ & $\Delta\kappa_{zz}$ & linear & $\Delta\kappa_{xx}$ & linear \\
\hline
crystal, 0.95 & $-0.36\%$ & $-0.12\%$ & $+0.97\%$ & $+1.11\%$ \\
crystal, 1.05 & $-0.12\%$ & $+0.12\%$ & $-1.21\%$ & $-1.06\%$ \\
crystal, 1.2 & $-2.86\%$ & $+0.44\%$ & $-5.69\%$ & $-3.96\%$ \\
crystal, 1.5 & $-12.3\%$ & $+0.98\%$ & $-14.6\%$ & $-8.80\%$ \\
amorphous, 0.95 & $+2.45\%$ & $+2.68\%$ & $+1.61\%$ & $+1.77\%$ \\
amorphous, 1.05 & $-2.71\%$ & $-2.55\%$ & $-1.80\%$ & $-1.68\%$ \\
amorphous, 1.2 & $-11.3\%$ & $-9.51\%$ & $-7.22\%$ & $-6.29\%$ \\
amorphous, 1.5 & $-26.1\%$ & $-21.2\%$ & $-15.5\%$ & $-14.0\%$ \\
\end{tabular}
\end{ruledtabular}
\end{table}

\section{Discussion}\label{sec:discussion}

\paragraph*{Two complementary local views.}
The sensitivity maps avoid the partition altogether, in the way forces do for the energy: the
energy of an atom depends on the gauge, but the derivative of the total energy with respect to its
position, the Hellmann--Feynman force, is unique. Here the share of $\kappa$ belonging to an atom is
ambiguous, but the derivative of $\kappa$ with respect to its mass or its bonds is not; ``alchemical''
derivatives of the energy with respect to nuclear charge follow the same
logic~\cite{Lilienfeld2009accurate}. Differentiable transport exists at the continuum level, through adjoint and
automatic-differentiation solvers of the phonon Boltzmann equation~\cite{Romano2022inverse,Shang2025jaxbte}, and bond-level
response has been used to tune elastic moduli~\cite{Goodrich2015principle}; interface conductance modal
analysis~\cite{Gordiz2015formalism} resolves the conductance of an interface over pairs of modes. To our knowledge,
none of these gives the exact response of an atomistic thermal conductivity to local masses and bonds. The heat-flow map answers ``where does heat flow?'': through
regions of small vibrational amplitude, that is, of low local compliance, whose bonds are on average slightly
longer; the low compliance is a property of the network, not of its individual bonds. This description is robust at the 4~\AA\ scale despite the gauge freedom of the local energy and
across the tested interatomic potentials. The sensitivity map answers ``where would a change matter?'' Its answer is largely
the complementary region: in soft, large-amplitude regions that carry little heat, $\km$ is large and
positive. The exact decomposition,
Eq.~(\ref{eq:decomp}), makes the link explicit. The current term ties $\km$ positively to $\zeta$; the
rotation term reverses the sign by redistributing sensitivity through the mixing of nearby modes; and the
local vibrational amplitude, a low-frequency moment of the same local mode weights, predicts where that
redistribution goes.

\paragraph*{Which map to use.} In our view the two maps answer different questions, at comparable cost when
computed densely (Sec.~\ref{sec:models}). The site projection also has an $O(N)$ route, through the Chebyshev expansion, for $\kappa$ and for any
fixed set of sites, which we used for BM10k; for the sensitivities we have not implemented one, but a route
exists: $\km$ is the gradient of a single scalar, so reverse-mode (adjoint) differentiation of the stochastic
Chebyshev trace for $\kappa$ yields the whole map at a cost comparable to a few evaluations of $\kappa$, with a
stochastic error at each site that we have not assessed. Meanwhile $\dR$, from sparse linear solves, is a cheap proxy. Thus the site projection is the tool for asking where heat flows: screening
large models, finding channels and rattlers, and transport along an interface or a filament, where the
current flows in parallel. It needs regional averages because of its gauge. The sensitivity maps are the
tool for asking what to change: design, single sites, and geometries in which the layers conduct in series,
such as the interface of Sec.~\ref{sec:interface}, where the flux is the same in every layer and the site
projection is nearly uniform whatever the resistance. The practice we would recommend is to use both: the
site projection first, to see where heat flows, and the sensitivity map where a decision about what to
change must be made.

\paragraph*{Relation to soft spots and elastic heterogeneity.}
The large-$\dR$ regions are the vibrational analogue of the soft spots of sheared amorphous
packings, identified from quasi-localized low-frequency modes~\cite{Manning2011soft,Xu2010anharmonic},
and of the regions of low local modulus that Mizuno, Mossa and Barrat linked to the vibrational
anomalies and the thermal conductivity of model glasses~\cite{Mizuno2013elastic,Mizuno2016elastic}.
Here the same regions are where little heat flows and where the conductivity is most easily changed.
Gautam~\emph{et al.} found, in five amorphous, paracrystalline and polycrystalline silicon models, that
sites with large thermal displacements (from molecular dynamics and in the harmonic approximation) have
low normalized SPTC, and described them as rattlers~\cite{Gautam2026lattice}. Our rank statistics,
block-bootstrap intervals and partial correlations, over six networks and three potentials, confirm this
relation and extend it from the heat-flow map to the response of $\kappa$.

\paragraph*{Limitations.}
The calculations are harmonic, at the $\Gamma$ point, in cells of up to 10\,000 atoms. (i) The
propagon contribution is cell-size limited and is not the subject of this work; it requires a treatment
beyond the harmonic AF picture~\cite{Simoncelli2019unified,Isaeva2019unified} or modal analysis of
molecular dynamics~\cite{Larkin2014accumulation,Lv2016gkma}; the diffuson conductivity
is converged (0.944--0.963~\WmK{} in the 4096-atom cell and in a 32\,768-atom replica; Supplemental
Material, Sec.~\textcolor{red}{S2}). (ii) The maps use a uniform broadening, and the physical lifetimes justify it for $\kappa$. Three-phonon
widths from finite-difference cubic force constants (median half-width 0.10~meV at 300~K) give
$\kappa=1.208$~\WmK{} for si4096 and 1.270~\WmK{} for the 32\,768-atom replica, within 3\% of the $\eta=2$~K values
(1.218 and 1.237). The maps are another matter. Many physical widths fall below the level spacing of a
4096-atom cell; the mass map then becomes large and noisy (the Euler rule still closes, to $10^{-7}$, but the
largest $|\km|$ is 7 times its value at $\eta=2$~K and 19\% of the sites exceed ten times the spread of the
$\eta=2$~K map), and the anti-correlation with $\zeta$ is lost: Spearman $-0.05$ at 4~\AA\ with raw widths and
$-0.03$ to $-0.17$ with widths floored at one to four local level spacings, against $-0.41$ at $\eta=2$~K
(Supplemental Material, Table~\textcolor{red}{S2}). The anti-correlation returns steadily as the floor rises, so
in our view this is a finite-cell effect, not physics, and it is reasonable to conjecture that a map with
physical widths needs cells in which the widths exceed the level spacing. Until then, region-averaged maps at $\eta=2$--10~K are the
practical choice, and the anti-correlation is a property of those maps. (iii) Mass maps from two potentials agree well above chance but only moderately: between ACE
and GAP-18 for si4096 the Spearman correlation is 0.47 at single sites and 0.51 at 4~\AA\ at $\eta=2$~K, and 0.63 and 0.71
at 10~K (shift nulls 0.05, 0.15, 0.03 and 0.08), against 0.89--0.94 for $\zeta$. The heat-flow map is thus more
transferable between potentials than the response map. (iv) A per-atom Green--Kubo check with fully anharmonic molecular dynamics (si512) supports the
rattler picture qualitatively (Supplemental Material, Sec.~\textcolor{red}{S17} and Table~\textcolor{red}{S11}).

\paragraph*{Bonds.}
The bond sensitivities (Supplemental Material, Table~\textcolor{red}{S12}) are dominated by bending for all three potentials, and
stiffening the bonds around heat-carrying atoms raises $\kappa$ in every case. Their dependence on bond
length is potential specific: for ACE and GAP-18 long bonds are the most sensitive, through stretching,
whereas the Keating form, whose stretching constant grows with bond length, reverses the trend.

\paragraph*{Design.}
For isotopes and small changes of bond stiffness the linear map is quantitative. For heavy substituents
it is a ranking tool when the substituents are dispersed, and it does not distinguish the compact-cluster
placements tested here. The
connection to $\dR$ also gives a cheap screening variable: $\dR$ needs only sparse linear solves and
predicts the sensitivity map with Spearman 0.3--0.9 at 4~\AA.

\section{Conclusions}

The site-projected thermal conductivity of a-Si depends moderately on the choice of local energy at single sites
but is robust over regions and across the tested potentials. Its gauge-free counterpart, the exact response of
$\kappa$ to local masses and bonds, obeys an Euler sum rule and contains the heat-flow map as one of three exact
terms. With ACE, GAP-18 and Keating force constants the two maps are anti-correlated; the anti-correlation strengthens with
cell size and when longer-wavelength modes are included, and at room
temperature the local vibrational amplitude largely explains it: soft regions carry little heat, and there a heavier
atom lowers $\kappa$ most, through the mixing of modes about 1--2~meV apart. In paracrystalline silicon the grains
conduct more heat per atom and the matrix controls the response. The response map guides the placement of dispersed
substituents and locates the region, on the amorphous side of an amorphous/crystalline interface, that controls
cross-plane transport.

More broadly, the gauge question that has shadowed local descriptions of heat transport has a constructive answer:
ask how $\kappa$ responds, not how it is shared. The construction needs only a harmonic network and applies to any
transport coefficient written as a trace over eigenstates, the Kubo--Greenwood electrical conductivity among them;
we expect the distinction between where a current flows and where it can be changed to be general. Natural
applications are the amorphous dielectrics and interfaces of nanoscale devices, and thermoelectrics, where the
electrical analogue would locate the regions that control both currents.

\begin{acknowledgments}
We thank Dr.~N.~Bernstein for providing the amorphous/crystalline interface model, and L.~A.~M.~Rosset and V.~L.~Deringer for the paracrystalline models R1--R3 and the SiH-ACE-25 potential. We thank A.~Gautam for
his earlier work on site-projected thermal conductivity and on rattling sites in silicon networks, on
which this study builds. We acknowledge support
from the U.S. National Science Foundation (Grant No.~MRI-2320493) and the U.S. Office of Naval Research
(Grant No.~N00014-23-1-2773). C.U. acknowledges support from the Laboratory Directed Research and
Development program of Los Alamos National Laboratory under the Director's Postdoctoral Fellowship Program
(Project No.~20240877PRD4). Los Alamos National Laboratory is operated by Triad National Security, LLC, for
the National Nuclear Security Administration of the U.S. Department of Energy (Contract
No.~89233218CNA000001).
Computations were performed on a single workstation. Calculations, analysis code and drafting of this
manuscript were assisted by Claude Code (Anthropic), an AI system, in accordance with the APS policy on the
use of AI tools; the authors take full responsibility for the content.
\end{acknowledgments}

\paragraph*{Data availability.}
The structures, dynamical matrices, sensitivity maps and analysis scripts supporting this study are
openly available in Zenodo at \url{https://doi.org/10.5281/zenodo.23620133}.

\appendix

\section{Remarks on the original formulation}\label{app:remarks}

The following points correct or qualify Refs.~\cite{Gautam2025siteprojected,Ugwumadu2026mapping}; none
changes their conclusions about $\kappa$.
\begin{enumerate}
\item \emph{Asymmetric weight.} The AF weight $C_i/\omega_i^2$ is not symmetric in $(i,j)$. This leaves $\kappa$
  unchanged but makes site values depend on whether a bond current is anchored at its row or its column atom;
  the symmetrized weight of Sec.~\ref{sec:theory} removes the dependence (site values change by about 0.5\%).
\item \emph{Locality.} The decay of the site-pair matrix $\Xi({\ell},{\ell'})$~\cite{Gautam2025siteprojected,Ugwumadu2026mapping,Ugwumadu2026seamless},
  whose row sums give $\zeta({\ell})$, is inherited largely from the range of the force constants and is a power
  law, so the analogy with electronic nearsightedness is weaker than stated. At small broadening the response
  is non-local over several nanometres (Supplemental Material, Sec.~\textcolor{red}{S6}).
\item \emph{Negative site values} at localized modes are at least partly a property of the gauge-dependent
  partition, not numerical artefacts.
\item \emph{Gauge.} The non-uniqueness noted in Ref.~\cite{Ugwumadu2026mapping} is quantified in
  Secs.~\ref{sec:gauge} and \ref{sec:robust} and complemented by the gauge-free maps of this work.
\end{enumerate}

\section{The Euler rule}\label{app:euler}

Under $m_{\ell}\to\mu m_{\ell}$ for all ${\ell}$, $D\to D/\mu$ and $\Gamma^\alpha\to\Gamma^\alpha/\mu$ while the
eigenvectors are unchanged, so $P^\alpha\to P^\alpha/\mu$. The AF weight has the form
\begin{equation}
    F_{ij}\propto\frac12\left[\frac{C(\omega_i)}{\omega_i^2}+\frac{C(\omega_j)}{\omega_j^2}\right]\,r(\omega_i,\omega_j)\,
L_\eta(\omega_i-\omega_j),
\end{equation} 
with $C$ the mode heat capacity, $r$ a dimensionless ratio and $L_\eta$ a
normalized Lorentzian. If $T$ and $\eta$ are scaled with the frequencies, $T\to T/\sqrt\mu$ and
$\eta\to\eta/\sqrt\mu$, then $C$ and $r$ are unchanged, $\omega^{-2}\to\mu\,\omega^{-2}$ and
$L_\eta\to\sqrt\mu\,L_\eta$, so $F\to\mu^{3/2}F$ and $\kappa\to\mu^{-1/2}\kappa$.
Differentiating $\kappa(\mu m,\eta/\sqrt\mu,T/\sqrt\mu)=\mu^{-1/2}\kappa(m,\eta,T)$ at $\mu=1$ gives
\begin{equation}
\sum_{\ell}m_{\ell}\frac{\partial\kappa}{\partial m_{\ell}}-\frac12\eta\frac{\partial\kappa}{\partial\eta}
-\frac12T\Big(\frac{\partial\kappa}{\partial T}\Big)_C=-\frac12\kappa,
\end{equation}
which with $\km=-2m_{\ell}\partial\kappa/\partial m_{\ell}$ is Eq.~(\ref{eq:euler}). Here the temperature derivative
acts through the heat capacities only. The analogous rule for bonds follows from $\Phi\to\nu\Phi$.

\section{Derivation of the gradient}\label{app:gradient}

Write Eq.~(\ref{eq:kappa}) as $\kappa=\sum_\alpha\sum_{ij}F_{ij}(P^\alpha_{ij})^2$ with $F_{ii}=0$ and
$P^\alpha=E^{\mathsf T}\Gamma^\alpha E$, and consider a symmetric perturbation $D\to D+\delta D$. Three
quantities change: the operators $\Gamma^\alpha$, the eigenvalues, and the eigenvectors.

\emph{Explicit part.} Since $\Gamma^\alpha=D\circ R^\alpha$ is linear in $D$,
$\delta P^\alpha=E^{\mathsf T}(\delta D\circ R^\alpha)E$ at fixed $E$, and
$\delta\kappa=2\sum_\alpha{\rm Tr}\big[(F\circ P^\alpha)^{\mathsf T}E^{\mathsf T}(\delta D\circ R^\alpha)E\big]$,
which gives the second term of Eq.~(\ref{eq:grad}).

\emph{Eigenvalues.} First-order perturbation theory gives $\delta\lambda_i=(E^{\mathsf T}\delta DE)_{ii}$,
so the eigenvalue part is $\sum_i\tilde G_{ii}(E^{\mathsf T}\delta DE)_{ii}$, with
$\tilde G_{ii}=\partial\kappa/\partial\lambda_i$ the derivative at fixed eigenvectors, acting on $F$ only.

\emph{Eigenvectors.} For non-degenerate $i\neq k$,
\begin{equation}
\delta E_{\cdot i}=\sum_{k\neq i}E_{\cdot k}V_{ki}/(\lambda_i-\lambda_k),
\end{equation}
with $V=E^{\mathsf T}\delta DE$.
Then 
\begin{equation}
\delta P^\alpha_{ij}=\sum_{k\neq i}\frac{V_{ki}}{\lambda_i-\lambda_k}P^\alpha_{kj}+
\sum_{k\neq j}\frac{V_{kj}}{\lambda_j-\lambda_k}P^\alpha_{ik}.
\end{equation}
Inserting into $\delta\kappa=2\sum F_{ij}P^\alpha_{ij}\delta P^\alpha_{ij}$, using the symmetry of $F$
and the antisymmetry of $P^\alpha$ (so that $P^\alpha_{ji}P^\alpha_{jk}=P^\alpha_{ij}P^\alpha_{kj}$) and collecting
the coefficient of $V_{ik}$, gives
$\sum_{i\neq k}\tilde G_{ik}V_{ik}$ with $\tilde G_{ik}=N_{ik}/(\lambda_i-\lambda_k)$ and $N_{ik}$ of
Eq.~(\ref{eq:Gt}). $N$ is antisymmetric, so $\tilde G$ is symmetric, and Eq.~(\ref{eq:grad}) follows. This is the Daleckii--Krein formula~\cite{Daleckii1965integration,Bhatia1997matrix}: the
divided differences $(F_{ij}-F_{kj})/(\lambda_i-\lambda_k)$ in $\tilde G_{ik}$ tend to $\partial_1F$ as $\lambda_k\to\lambda_i$, which gives the near-degenerate form quoted in the main text. 

\emph{Mass sensitivity.} With $\delta D=-\tfrac12\delta\ln m_{\ell}\,(\Pi_{\ell}D+D\Pi_{\ell})$, where $\Pi_{\ell}$ projects on the coordinates of atom ${\ell}$, $s_m({\ell})=\sum_{a\in {\ell},b}G_{ab}D_{ab}$. The explicit part contracted with $D$ gives 
\begin{equation}
    2\zeta({\ell}) = 2\sum_\alpha\sum_{a\in {\ell}}\sum_bD_{ab}R^\alpha_{ab}[E(F\circ P^\alpha)E^{\mathsf T}]_{ab},
\end{equation}
and for the spectral part, 
\begin{equation}
\sum_{a\in {\ell}}[E\tilde GE^{\mathsf T}D]_{aa}=\sum_{ik}\tilde G_{ik}\lambda_k
c_{ik}({\ell}).
\end{equation}
The diagonal ($i=k$) gives $\seig({\ell})$; symmetrizing the off-diagonal sum over
$i\leftrightarrow k$ gives 
\begin{equation}
\srot({\ell}) = \sum_{i<k}\frac{N_{ik}(\lambda_i+\lambda_k)}{(\lambda_i-\lambda_k)\,c_{ik}({\ell})}.
\end{equation}
Hence $s_m=2\zeta+\seig+\srot$, and $\km=2s_m$ is Eq.~(\ref{eq:decomp}).

\bibliography{refs}

\end{document}


\title{Supplemental Material for ``Where heat flows and where to change it: gauge-free sensitivity maps of thermal conductivity in disordered solids''}
\author{D. A. Drabold}
\email{drabold@ohio.edu}
\affiliation{Department of Physics and Astronomy, Ohio University, Athens, Ohio 45701, USA}
\author{C. Ugwumadu}
\affiliation{Theoretical Division (T-4), Los Alamos National Laboratory, Los Alamos, New Mexico 87545, USA}

\maketitle

\noindent This Supplemental Material collects the checks, the secondary results and the ``nuts and bolts'' that support the main text but would interrupt it. The sections appear in the order in which the main text first calls on them. Most are short, and each says at the end what we take from it. A reader who wants only the physics may skip Secs.~\ref{sec:bondcomp}, \ref{sec:validation}, \ref{sec:potentials} and \ref{sec:on}, which concern definitions, numerics and models. Validation tests are collected in Sec.~\ref{sec:validation}.

\section{Bond compensation}\label{sec:bondcomp}
Scaling one bond of a many-body potential raises a small but real question of definition: the off-diagonal blocks $\Phi_{{\ell}{\ell'}}$ are not symmetric, so no change of a single bond keeps both the symmetry of the dynamical matrix and the acoustic sum rule exactly (Sec.~\textcolor{red}{II} of the main text). Table~\ref{tab:bondcomp} compares our choice with the obvious alternative.
\begin{table}[h]
\caption{Bond compensation. Our definition compensates the self blocks by $-\epsilon\,{\rm sym}(\Phi_{{\ell}{\ell'}})$; the
alternative scales ${\rm sym}(\Phi_{{\ell}{\ell'}})$ in all four blocks and preserves both the symmetry and the acoustic
sum rule bond by bond. Since $G_{{\ell}{\ell}}$ is symmetric, the two differ exactly by
$2\langle{\rm asym}\,G_{{\ell}{\ell'}},{\rm asym}\,\Phi_{{\ell}{\ell'}}\rangle$. Columns: relative size of ${\rm asym}(\Phi_{{\ell}{\ell'}})$ (median /
90th percentile, \%), Spearman correlation of the two maps, median relative change (\%), sign changes (\%), all
for nearest-neighbour bonds ($d<2.8$~\AA); overlap of the 100 largest bonds; relative Euler residual of our
definition and of the alternative.}
\label{tab:bondcomp}
\begin{ruledtabular}
\begin{tabular}{lccccccc}
model & asym (\%) & Spearman & change (\%) & sign (\%) & top 100 & Euler, ours & Euler, alt. \\
\hline
si4096 ACE 2K & 1.6 / 3.9 & 0.986 & 6.1 & 2.9 & 0.91 & $1\times10^{-6}$ & 22\% \\
si4096 ACE 10K & 1.6 / 3.9 & 0.996 & 3.0 & 1.5 & 0.97 & $1\times10^{-6}$ & 4\% \\
si4096 Keating 2K & 3.6 / 6.8 & 0.998 & 4.1 & 1.5 & 0.95 & $8\times10^{-7}$ & 15\% \\
R1 ACE 2K & 0.1 / 0.5 & 0.998 & 0.1 & 0.3 & 0.93 & $1\times10^{-6}$ & 10\% \\
R2 ACE 2K & 1.1 / 2.8 & 0.994 & 2.9 & 1.5 & 0.98 & $1\times10^{-6}$ & 13\% \\
R3 ACE 2K & 1.0 / 2.5 & 0.991 & 2.8 & 1.2 & 0.94 & $1\times10^{-6}$ & 13\% \\
si512 GAP 10K & 2.4 / 5.1 & 0.995 & 4.9 & 2.6 & 0.99 & $6\times10^{-6}$ & 7\% \\
\end{tabular}
\end{ruledtabular}
\end{table}

The two definitions give nearly the same bond maps, but only ours satisfies the Euler rule. Thus the choice matters for the sum rule and hardly at all for the physics.

\section{Broadening, widths and heat capacities}\label{sec:widths}
The broadening $\eta$ stands in for the true widths of the modes, and it is fair to ask what happens if one uses the true widths instead. Table~\ref{tab:widths} gives the answer for si4096.
\begin{table}[h]
\caption{si4096 (ACE): conductivity, Euler closure and mass-map statistics for different broadenings and
width models ($T=300$~K, quantum heat capacities unless noted). Floor: widths
$\gamma_i=\max[\gamma_{\rm anh}(\omega_i),c\,\Delta_{\rm loc}(\omega_i)]$ with $\Delta_{\rm loc}$ the local
level spacing, the mean spacing of the 20 modes nearest in frequency (ten on each side in the sorted
spectrum). Anharmonic widths are extrapolated as $\gamma\propto\omega^{1.53}$ below the lowest resolved mode;
with any floor the floor sets the widths there, so the floored rows do not depend on the extrapolation.
Correlations are with $\zeta$ from the same run (rows marked $^*$: $\zeta$ at $\eta=2$~K); the 95\%
block-bootstrap intervals of the Spearman values are $\pm0.03$--0.05 at single sites and $\pm0.05$--0.09 at
4~\AA. Last column: percentage of sites with $|\km|$ above ten times sd$(\km)$ of the $\eta=2$~K map. The raw-width
$\kappa$ (1.205) comes from the map code; the conductivity code used below gives 1.208.}
\label{tab:widths}
\begin{ruledtabular}
\begin{tabular}{lccccccc}
widths & $\kappa$ (\WmK{}) & Euler & sd$(\km)$ & max$|\km|$ & Spearman site / 4~\AA & Pearson site / 4~\AA & $>10$\,sd (\%) \\
\hline
$\eta=2$~K & 1.218 & $1\times10^{-6}$ & $9.8\times10^{-4}$ & 0.024 & $-0.31$ / $-0.41$ & $-0.31$ / $-0.44$ & 0.0 \\
$\eta=5$~K$^*$ & 1.067 & $3\times10^{-7}$ & $4.9\times10^{-4}$ & 0.003 & $-0.34$ / $-0.48$ & $-0.33$ / $-0.51$ & 0.0 \\
$\eta=10$~K & 0.948 & $6\times10^{-8}$ & $2.9\times10^{-4}$ & 0.002 & $-0.31$ / $-0.41$ & $-0.29$ / $-0.43$ & 0.0 \\
$\eta=20$~K$^*$ & 0.844 & $1\times10^{-8}$ & $1.5\times10^{-4}$ & 0.001 & $-0.16$ / $-0.19$ & $-0.14$ / $-0.20$ & 0.0 \\
$\eta=2$~K, classical $C$$^*$ & 1.331 & $1\times10^{-6}$ & $1.1\times10^{-3}$ & 0.025 & $-0.26$ / $-0.38$ & $-0.27$ / $-0.40$ & 0.0 \\
anharmonic, raw & 1.205 & $1\times10^{-7}$ & $8.4\times10^{-3}$ & 0.182 & $-0.04$ / $-0.05$ & $+0.01$ / $-0.02$ & 19.0 \\
floor $c=1$ & 1.262 & $2\times10^{-8}$ & $3.2\times10^{-3}$ & 0.098 & $-0.07$ / $-0.03$ & $-0.06$ / $-0.03$ & 0.3 \\
floor $c=2$ & 1.241 & $2\times10^{-8}$ & $2.3\times10^{-3}$ & 0.058 & $-0.10$ / $-0.09$ & $-0.12$ / $-0.09$ & 0.1 \\
floor $c=4$ & 1.199 & $1\times10^{-8}$ & $1.7\times10^{-3}$ & 0.015 & $-0.14$ / $-0.17$ & $-0.15$ / $-0.16$ & 0.1 \\
\end{tabular}
\end{ruledtabular}
\end{table}

Three-phonon widths were obtained from finite-difference cubic force constants of si512 and Fermi's golden
rule (median HWHM 0.10~meV at 300~K), extrapolated as $\gamma\propto\omega^{1.53}$ below the lowest resolved
mode. Without a floor, many widths fall below the local level spacing, and the mass map becomes large and
noisy: the Euler rule still closes, but sd$(\km)$ is 9 times its $\eta=2$~K value and 19\% of the sites lie beyond
ten times that spread. A floor of a few times the spacing shrinks the map and brings back part of the
anti-correlation with $\zeta$. The lesson is simple: a mode narrower than the level spacing of the cell is not resolved by the cell, and the map notices.

\paragraph*{Diffusons and propagons.} A 32\,768-atom replica ($2\times2\times2$ copies of si4096,
eight real $k$-points) with force constants truncated at 11~\AA\ and the sum rule restored gives the
diffuson conductivity (10--56.5~meV) 0.944~\WmK{}, compared with 0.948--0.963 in the 4096-atom cell for all width
models. The diffuson part is converged. The propagon part ($<10$~meV) is not. With physical widths
(three-phonon widths extrapolated as $\gamma\propto\omega^{1.53}$ below the lowest resolved mode) it is
0.241~\WmK{} (4096; equal, by coincidence, to the $\eta=2$~K value) and 0.306~\WmK{} (replica), and total
$\kappa$ is 1.208 and 1.270~\WmK{}, within 3\% of the $\eta=2$~K values (1.218, 1.237). Three-phonon widths are resolved only down to the lowest mode of si512 (about 7.5~meV) and must be
extrapolated below it. An earlier extrapolation held them constant there, which over-damps the lowest modes
(the physical widths fall roughly as $\omega^{1.5}$). In a periodic cell larger widths \emph{raise} the AF
propagon value, so that extrapolation gave propagon contributions that are too high, 0.310 (4096) and
0.402~\WmK{} (replica); it is superseded by the $\omega^{1.53}$ extrapolation quoted above. In a periodic cell the AF propagon conductivity decreases when the widths
fall below the level spacing, so finite cells cannot represent long-mean-free-path propagons. Thus the diffusons, which dominate $\kappa$ and the maps, are in good hands; the propagons are not, and we make no claim about them.

\section{Validation}\label{sec:validation}
An exact gradient is only as good as its implementation. The checks below all pass by wide margins.
\begin{itemize}
\item Analytic gradient against central finite differences from an independent code (si512, ACE,
$\eta=2$~K): median relative error $2\times10^{-5}$ over all 512 mass sensitivities, and
$10^{-6}$--$10^{-5}$ over 48 bond sensitivities. The bond check requires the symmetrized self-block
compensation ${\rm sym}(\Phi_{{\ell}{\ell'}})$, because the off-diagonal blocks of a many-body potential are not
symmetric.
\item Euler rule, Eq.~(\textcolor{red}{6}) of the main text: closes to $10^{-6}$ in every $\eta$-broadened run (all models,
potentials, temperatures and heat-capacity models), and to $10^{-7}$ with physical widths, raw or floored
(Table~\ref{tab:widths}).
\item Decomposition: $\sum_{\ell}s_\Gamma=2\kappa$ to round-off; the three terms reconstruct $\km$ to
$10^{-16}$; $\sum_{\ell}\srot/\kappa\le10^{-15}$ in every run; $\sum_{\ell}2\seig/\kappa=-2.86$ to $-3.11$.
\item The compact pair formula for $\srot$ [Eq.~(\textcolor{red}{10}) of the main text], with $N_{ik}$ built directly from
the eigenvectors, reproduces the code to a maximum difference of $8.7\times10^{-15}$~\WmK{}, against a largest
$|\srot({\ell})|$ of $2.1\times10^{-2}$~\WmK{} (si512, $\eta=2$~K). $N$ is antisymmetric to round-off.
\item Conjugate-gradient $\dR$ against the dense pseudo-inverse of the same sparse matrix: maximum
relative difference $1.9\times10^{-6}$ (si512) and $1.0\times10^{-6}$ (si4096); truncating the matrix at
11~\AA\ changes $\dR$ by at most 0.4\%.
\end{itemize}

\section{Potentials and structures}\label{sec:potentials}
We use three descriptions of the forces on purpose: two machine-learned potentials of near first-principles quality, and the Keating model, which is simple, transparent and in some respects wrong. A result that survives all three is unlikely to be an artifact of one. Table~\ref{tab:models} lists the models.
\begin{table}[h]
\caption{Models. Coordination defects counted with a 2.8~\AA\ cutoff after relaxation with the potential
used for the maps. $^a$A 216-site diamond cell ($a=4\times2.35/\sqrt3$~\AA) with 20 sites removed at random and
Gaussian displacements of 0.3~\AA, relaxed with the original Stillinger--Weber parameters; used only for the exact gauge
comparisons of Sec.~\textcolor{red}{II~B} of the main text.}
\label{tab:models}
\begin{ruledtabular}
\begin{tabular}{lllll}
model & $N$ & relaxation & 3/5/6-fold & defects \\
\hline
si4096 (DTW~\cite{Djordjevic1995computer}), ACE & 4096 & fixed cell (43.42~\AA) & 5/23/0 & 0.68\% \\
si4096, Keating & 4096 & fixed cell, bond topology of the input & (all four-bonded) & --- \\
si4096, GAP-18 & 4096 & fixed cell (43.42~\AA) & 2/38/1 & 1.0\% \\
si512, ACE and GAP-18 & 512 & fixed cell & 0/0/0 & 0 \\
BM10k relaxed & 10\,000 & ACE, zero pressure & 16/36/4 & 0.56\% \\
BM10k, 800~K anneal & 10\,000 & ACE MD, then zero pressure & 51/73/0 & 1.24\% \\
BM10k, 1000~K anneal & 10\,000 & ACE MD, then zero pressure & 63/87/0 (one 2-fold) & 1.51\% \\
R1 / R2 / R3 & 1000 & ACE, fixed cell & 10/0/0, 5/3/0, 3/3/0 & 1.0, 0.8, 0.6\% \\
interface & 2022 & GAP, then ACE & 19/31/0 (all amorphous) & 2.5\% (3.6\% of a-Si) \\
SW gauge tests & 196 & SW, fixed cell (16.28~\AA)$^a$ & 33/1/0 (one 2-fold) & 18\% \\
\end{tabular}
\end{ruledtabular}
\end{table}

\paragraph*{Keating.} $E=\sum_{\rm bonds}\frac{3\alpha}{16d^2}(r^2-d^2)^2+
\sum_{\rm angles}\frac{3\beta}{8d^2}(\bm r\cdot\bm r'+d^2/3)^2$ with $\alpha=2.965$~eV/\AA$^2$
(47.5~N/m), $\beta=0.285\alpha$ (13.5~N/m), $d=2.35$~\AA, on the bond list of the input network. With a
2.8~\AA\ distance criterion the relaxed Keating structure shows 16 five- and 2 six-fold atoms, but these
are geometric contacts; the energy keeps the four-bonded topology. The stretching constant of this form,
$\propto3r^2-d^2$, rises with bond length, whereas Si bonds soften when stretched; this is why the bond
sensitivity of Keating has the opposite length dependence to ACE and GAP-18 (Sec.~\ref{sec:bonds}).

\paragraph*{GAP-18.} The \texttt{gp\_iter6\_sparse9k.xml} potential~\cite{Bartok2018silicon}, run in LAMMPS
through QUIP. Diamond Si has its minimum at $a\approx5.46$~\AA; per-atom energies sum to the total. For si512
GAP-18 gives $\kappa=0.944$~\WmK{} (ACE 1.091) at $\eta=2$~K and an optic peak at 57.5~meV (ACE 59.8~meV). For
si4096 the force constants come from sparse finite differences with graph colouring, truncated at 10~\AA,
symmetrized and with the acoustic sum rule restored; applied to si512, the same procedure reproduces the dense
matrix to $5\times10^{-8}$ (largest relative difference of a block; asymmetry before symmetrization
$2.6\times10^{-3}$). It gives $\kappa=1.009$~\WmK{} (ACE 1.218) at $\eta=2$~K and 0.867 (ACE 0.948) at 10~K.

\section{Vibrational densities of states}\label{sec:vdos}
\begin{figure}[h]
\includegraphics[width=\textwidth]{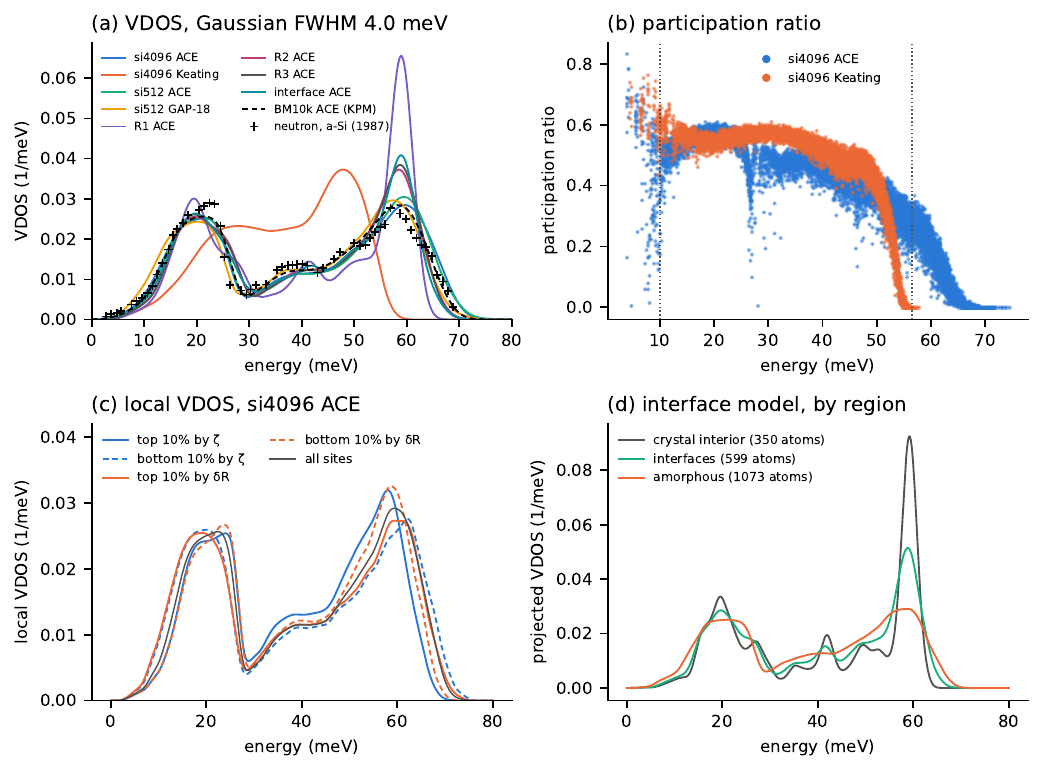}
\caption{(a) Vibrational density of states of all models, Gaussian broadened (FWHM 4~meV), against the
inelastic neutron data for a-Si of Kamitakahara \emph{et al.}~\cite{Kamitakahara1987vibrational} (crosses).
BM10k from a kernel-polynomial expansion (1500 moments, 32 random vectors) of its sparse ACE dynamical
matrix. (b) Participation ratio against energy, si4096 (ACE and Keating); dotted lines at 10~meV and at
56.5~meV, the boundaries used for propagons and locons. (c) Local VDOS of si4096 (ACE) averaged over the
10\% of sites with the largest and smallest $\zeta$ and $\dR$ (4~\AA\ averages), and over all sites. Every
curve in all panels is normalized to unit area over 0--80~meV, the neutron data over their measured range;
panels (c) and (d) therefore compare shapes, not the numbers of atoms in each group.
(d) Interface model: VDOS projected on the crystal more than 4~\AA\ from the interfaces, on atoms within
4~\AA\ of the two interfaces, and on the amorphous region more than 4~\AA\ from them.}
\label{fig:vdos}
\end{figure}

\paragraph*{Experimental comparison.} The measured spectrum was digitized from Fig.~3(a) of
Ref.~\cite{Kamitakahara1987vibrational} (58 data points, the crosses; the energy axis was calibrated on the
five energy values marking the cuts in that figure, with residuals below 0.13~meV). The reference does not report the instrumental
resolution (time-of-flight data on IN4 with incident energy 12~meV, so that most of the spectrum is
measured in neutron energy gain, where the resolution varies with energy), so we broadened the
calculations with a Gaussian whose width minimizes the rms difference from the data; this constant width
is an effective resolution: FWHM 4.5~meV for ACE (si4096) and
3.5~meV for GAP-18 (si512), with misfits of 9 and 10\% of the peak height. Panel (a) uses 4~meV for all
models. ACE and GAP-18 reproduce the TA-like band, the minimum near 30~meV and the TO-like band; the
paracrystalline models R1--R3 have sharper peaks than the amorphous measurement. The Keating spectrum has
its TO-like peak near 48~meV (ACE: 59.6~meV), about 10~meV below the measured band, and its TA-like
maximum near 28~meV (ACE: 21.5~meV); its TA-like band extends too high and fills the minimum near 30~meV
($g$ at 30~meV is 0.99 of the TA-like maximum, against 0.22 for ACE). Kamitakahara \emph{et al.} found
the TA-like band of their own Keating model too high as well (26 against 20~meV).

\paragraph*{Local spectra.} The weight below 12.4~meV is 0.034 averaged over all sites, 0.027 and 0.051
for the 10\% of sites with the largest and smallest $\zeta$, and 0.054 and 0.027 for the 10\% with the
largest and smallest $\dR$. High-$\zeta$ sites also have a slightly softer optic band (mean frequency 39.8
against 40.4~meV for low-$\zeta$ sites; weight above 50~meV 0.38 against 0.41), consistent with their longer bonds. Thus the regions that carry heat are, in a modest way, spectrally different from those that do not, which is the local softness of the main text seen from the side of the spectrum.

\section{O(N) evaluation}\label{sec:on}
Dense diagonalization stops at a few thousand atoms. For larger systems one needs a route that scales linearly with $N$, and this section describes the one we used for BM10k. Writing Eq.~(\textcolor{red}{3}) of the main text as the diagonal of
$\sum_{nm}c_{nm}T_n(\tilde D)\Gamma^\alpha T_m(\tilde D)\Gamma^{\alpha\mathsf T}$, with Chebyshev
polynomials of the rescaled $\tilde D$, two-dimensional Chebyshev coefficients $c_{nm}$ of the AF weight
and no damping kernel, reproduces $\kappa$ with a relative error of
0.03\% and $\zeta$ to 0.9\% of its spread with $N_c=1000$
(si4096, $\eta=2$~K); the three acoustic zero modes are projected out exactly. $\kappa$ is computed as a
stochastic trace (32 random-sign vectors). Per-site $\zeta$ would need $\sim10^5$ random vectors and is
computed instead with unit vectors on each requested site. The cost is therefore $O(N)$ for $\kappa$ and for a
fixed number of sites (each site costs $O(N N_c)$), but $O(N^2N_c)$ for the complete map of all sites. The GPU implementation (CuPy) has memory linear
in $N$ and reproduces exact results at 4096 atoms ($\zeta$: Pearson 0.9999, rms error 1.5\% of the spread)
and on the 32\,768-atom replica ($\kappa$ $1.227\pm0.008$ against 1.237). $\zeta$ is not local at
small broadening: spheres of 25~\AA\ reproduce 93\% of its spatial scale at $\eta=10$~K but only 40\% at
$\eta=2$~K, so spatial partitioning is not a route to large systems. In plain terms: at small broadening the heat flow at a site knows about the structure several nanometres away.

The sparse ACE dynamical matrix of BM10k was obtained from LAMMPS forces with graph colouring: atoms
farther apart than $2r_c+1$~\AA\ are displaced in the same configuration, and blocks $D_{kj}$ are kept for
$|R_k-R_j|<11$~\AA, with the acoustic sum rule restored on the self blocks. It reproduces the dense
matrices of si512 and si4096 to $10^{-8}$ and $3\times10^{-6}$.

\paragraph*{Defects in BM10k.} The three- and five-fold sites of BM10k are soft: $\dR=0.18$--0.20~\AA\ at three-fold
and 0.16~\AA\ at five-fold sites, against 0.135--0.137~\AA\ at four-fold sites ($p<10^{-10}$), and they carry 13--37\%
less heat. Restricted to four-fold sites, $S(\zeta,\dR)$ is still $-0.47$ to $-0.49$ (Sec.~\textcolor{red}{IV~C} of the main text), so the
relation is not carried by the defects.

\section{Local softness as a diagonal of the pseudo-inverse}\label{sec:dRinv}

In the classical harmonic approximation a mass-weighted normal coordinate has
$\langle q_i^2\rangle=k_BT/\lambda_i$, and the Cartesian displacement of coordinate $a$ is
$u_a=m_{{\ell}(a)}^{-1/2}\sum_iE_{ai}q_i$. Hence
$\langle u_a^2\rangle=(k_BT/m_{{\ell}(a)})\sum_iE_{ai}^2/\lambda_i=(k_BT/m_{{\ell}(a)})[D^+]_{aa}$, where the sum
and the pseudo-inverse exclude the three uniform translations. Summing over the three coordinates of ${\ell}$
gives Eq.~(\textcolor{red}{11}) of the main text. To evaluate $[D^+]_{aa}$ without eigenvectors we solve $Dy=Qe_a$, where
$Q=1-\sum_tt\,t^{\mathsf T}$ removes the mass-weighted translations $t$, by conjugate gradients with a
block-Jacobi preconditioner (the inverse of each atom's $3\times3$ diagonal block), reapplying $Q$ at every
step so that $y=D^+Qe_a$ and $[D^+]_{aa}=e_a^{\mathsf T}y$. A relative residual of $10^{-12}$ takes about
210 iterations for si4096. The quantum generalization replaces $k_BT/\lambda_i$ by
$(\hbar/2\omega_i)\coth(\hbar\omega_i/2k_BT)$ and requires the spectrum.

\section{Site currents in an arbitrary gauge}\label{sec:sitecurrents}

With $u=M^{-1/2}\sum_iE_iq_i$ and $\dot u=M^{-1/2}\sum_iE_i\dot q_i$ in
$\mathbf J=\sum_{\ell}\mathbf j_{\ell}$, and $q_i,\dot q_i$ written with creation and annihilation operators, the
coefficient of $a_i^\dagger a_j$ is proportional to $(B_{ji}\omega_i-B_{ij}\omega_j)/\sqrt{\omega_i\omega_j}$.
Normalizing to the AF form $\propto P_{ij}(\omega_i+\omega_j)/\sqrt{\omega_i\omega_j}$ gives
Eq.~(\textcolor{red}{4}) of the main text; the terms $a_ia_j$ and $a_i^\dagger a_j^\dagger$ do not contribute at zero frequency.
For the AF gauge, $\tilde{\mathbf A}_{mn}=\tfrac12(\mathbf R_m-\mathbf R_n)D_{mn}$ is antisymmetric and
$P^{\rm eff}=P$. The site-gauge map replaces one factor $P$ in Eq.~(\textcolor{red}{1}) of the main text by the current of site
${\ell}$ alone, built from $\mathbf A^{({\ell})}$; its evaluation with per-atom energies only is described in
Sec.~\ref{sec:gaugeSM}.

\section{Gauge dependence of the site projection}\label{sec:gaugeSM}
To measure the gauge dependence of $\zeta$ we need the map in a second gauge, and the natural second gauge is the one the potential itself supplies through its atomic energies. Here we describe how we computed it without ever forming the per-atom Hessians. In the site-energy gauge the local energy is $h_{\ell}=p_{\ell}^2/2m_{\ell}+\tfrac12u^{\mathsf T}H^{({\ell})}u$ with
$H^{({\ell})}=\partial^2\varepsilon_{\ell}/\partial u\,\partial u$, where $\varepsilon_{\ell}$ is the energy the potential
assigns to atom ${\ell}$. Then
\begin{equation}
\zeta_{\rm site}({\ell})=\sum_{mn}H^{({\ell})}_{mn}S^{({\ell})}_{mn},\qquad
S^{({\ell})}_{mn}=\sum_\alpha(R_{\ell}-R_n)^\alpha Y^\alpha_{mn}/\sqrt{m_mm_n},
\end{equation}
where $S^{({\ell})}$ is symmetrized, $S^{({\ell})}\to\tfrac12(S^{({\ell})}+S^{({\ell})\mathsf T})$, because only its symmetric part
contributes to the contraction with the symmetric $H^{({\ell})}$, and $Y^\alpha=EZ^\alpha E^{\mathsf T}$ with
$Z^\alpha=F\circ P^\alpha$, the matrix of Eq.~(\textcolor{red}{3}) of the main text.
$Y$ is taken from the total (AF) current: the gauge changes the total current only off shell, which costs
an $O(\eta)$ error in $\zeta$. $H^{({\ell})}$ is never formed. We diagonalize the small matrix
$S^{({\ell})}=\sum_j\mu_jv_jv_j^{\mathsf T}$ (with dimension equal to three times the number of atoms within the
site-energy cutoff of ${\ell}$), and evaluate $\zeta_{\rm site}({\ell})=\sum_j\mu_jv_j^{\mathsf T}H^{({\ell})}v_j$ with
$v^{\mathsf T}H^{({\ell})}v=[\varepsilon_{\ell}(R+hv)+\varepsilon_{\ell}(R-hv)-2\varepsilon_{\ell}(R)]/h^2$, which needs only
per-atom energies; the displacement step is $h=0.005$~\AA. Sites farther apart than twice the cutoff are probed in the same configuration. For
Stillinger--Weber, where the exact site-gauge result is available, the probed $\zeta$ agrees with it to
0.3\%.

For si4096 with ACE, the AF and site-energy maps have Pearson and Spearman correlation 0.92 at single
sites; 819 of the 1024 top-quartile sites and 828 of the 1024 bottom-quartile sites agree in both gauges.
The two maps differ by a uniform offset: $\kappa$ in the site-energy gauge is 0.78\% lower than in the AF gauge
($-2.3\times10^{-6}$~\WmK{} per site), the $O(\eta)$ difference of result (i) of the main text. No averaging removes it,
and left in it makes the ratio below rise again beyond 6~\AA\ (0.157 at 8~\AA, 0.192 at 10~\AA). It is a global
difference, not a local one, so we quote it once and remove it. The local gauge uncertainty (standard deviation over
gauges, offset removed, relative to the spatial spread of the mean) is then 0.227 at single sites, 0.124 at 2.6~\AA,
0.095 at 4~\AA, 0.077 at 5~\AA, 0.070 at 6~\AA, 0.055 at 8~\AA\ and 0.046 at 10~\AA\ (Spearman 0.917, 0.975,
0.984, 0.989, 0.990, 0.993, 0.995). Beyond 4~\AA\ it is fitted well by $0.015+0.32~\text{\AA}/R$; it halves between 4 and
10~\AA, against a factor 2.5 for pure $1/R$. For the interface model it is 0.04. Thus the local gauge ambiguity falls roughly as $r_s/R$
beyond about 4~\AA\ ($r_s=6.4$~\AA\ for ACE), as expected for a surface term, and the ranking of sites is nearly gauge-invariant.

\section{Anti-correlation and mediation at $\eta=10$~K}\label{sec:eta10}
The anti-correlation and its mediation by local softness do not depend on the choice of broadening. Table~\ref{tab:eta10} repeats the three si4096 rows of Table~\textcolor{red}{II} of the main text at $\eta=10$~K; the Keating anti-correlation is, if anything, stronger.
\begin{table}[H]
\caption{As Table~\textcolor{red}{II} of the main text, at $\eta=10$~K. $S(x,y)$: Spearman rank correlation of maps $x$ and $y$ (4~\AA\ averages, 95\% block-bootstrap intervals); $S(\km,\zeta\,|\,\dR)$: partial correlation of $\km$ and $\zeta$ at fixed $\dR$.}
\label{tab:eta10}
\begin{ruledtabular}
\begin{tabular}{lccccc}
model, potential, $\eta$ & $\kappa$ (\WmK{}) & $S(\km,\zeta)$ & $S(\zeta,\dR)$ & $S(\km,\dR)$ & $S(\km,\zeta\,|\,\dR)$ \\
\hline
si4096, ACE, 10~K & 0.948 & $-0.41$ $[-0.46,-0.36]$ & $-0.70$ $[-0.73,-0.66]$ & $+0.68$ $[+0.64,+0.72]$ & $+0.12$ $[+0.06,+0.18]$ \\
si4096, Keating, 10~K & 1.166 & $-0.85$ $[-0.87,-0.83]$ & $-0.90$ $[-0.91,-0.88]$ & $+0.90$ $[+0.89,+0.91]$ & $-0.23$ $[-0.32,-0.16]$ \\
si4096, GAP-18, 10~K & 0.867 & $-0.22$ $[-0.27,-0.16]$ & $-0.61$ $[-0.66,-0.56]$ & $+0.55$ $[+0.50,+0.59]$ & $+0.18$ $[+0.12,+0.25]$ \\
\end{tabular}
\end{ruledtabular}
\end{table}

\section{Sign and magnitude of the mass map}\label{sec:sign}
Since $\km$ takes both signs, ``anti-correlated with the heat flow'' could mean two things: that the \emph{sign} of the response follows the heat flow, or that its \emph{size} does. Table~\ref{tab:sign} separates the two. The sign follows reliably; the size follows only on average and only after spatial averaging.
\begin{table}[h]
\caption{Signed and absolute mass sensitivity against the heat-flow map: fraction of sites with $\km<0$;
Spearman correlation of $\zeta$ with $\km$ and with $|\km|$ at single sites and over 4~\AA\ spheres, and with the
absolute value of the 4~\AA\ average of $\km$; ratio of the mean $|\km|$ of the 10\% of 4~\AA\ regions with the
smallest $\zeta$ to that of the 10\% with the largest. 95\% block-bootstrap intervals are $\pm0.02$--0.20 (Sec.~\textcolor{red}{III} of the main text gives the block counts).}
\label{tab:sign}
\begin{ruledtabular}
\begin{tabular}{lccccccc}
model & $\km<0$ & site: $\km$ & site: $|\km|$ & 4~\AA: $\km$ & 4~\AA: $|\km|$ & $|\langle\km\rangle_{4\,\text{\AA}}|$ & $|\km|$ ratio \\
\hline
si4096 ACE 2K & 0.38 & $-0.31$ & $-0.02$ & $-0.41$ & $-0.18$ & $-0.30$ & 2.3 \\
si4096 ACE 10K & 0.20 & $-0.31$ & $-0.20$ & $-0.41$ & $-0.32$ & $-0.41$ & 1.8 \\
si4096 Keating 2K & 0.42 & $-0.63$ & $-0.24$ & $-0.66$ & $-0.41$ & $-0.39$ & 2.4 \\
R1 ACE 2K & 0.44 & $-0.27$ & $-0.22$ & $-0.51$ & $-0.67$ & $-0.43$ & 6.1 \\
R2 ACE 2K & 0.43 & $-0.31$ & $+0.06$ & $-0.55$ & $-0.03$ & $-0.16$ & 1.3 \\
R3 ACE 2K & 0.40 & $-0.29$ & $-0.08$ & $-0.43$ & $-0.31$ & $-0.14$ & 1.3 \\
si512 GAP 10K & 0.27 & $-0.18$ & $-0.01$ & $-0.00$ & $+0.18$ & $-0.00$ & 1.0 \\
si4096 GAP 2K & 0.36 & $-0.28$ & $-0.02$ & $-0.31$ & $-0.14$ & $-0.25$ & 2.0 \\
si4096 GAP 10K & 0.23 & $-0.28$ & $-0.13$ & $-0.22$ & $-0.09$ & $-0.22$ & 1.4 \\
\end{tabular}
\end{ruledtabular}
\end{table}

\section{Decomposition of the covariance and pair-resolved rotation term}\label{sec:cov}
The exact decomposition $\km=4\zeta+2\seig+2\srot$ lets us ask which term carries the anti-correlation and, for the rotation term, which pairs of modes carry it. Table~\ref{tab:cov} gives the answer: the rotation term, and pairs of modes a few $\eta$ apart in the 10--30~meV band.
\begin{table}[h]
\caption{Top: share of the covariance between $\km$ and $\zeta$ (4~\AA\ averages) carried by each term
of $\km=4\zeta+2\seig+2\srot$, for every model. Bottom: share of cov$(2\srot,\zeta)$ for si4096 by mode-pair separation (in units of
$\eta$) and by the \emph{mean frequency} $(\omega_i+\omega_k)/2$ of the pair ($\eta=2$~K unless noted).
$^a$The total covariance is close to zero (Spearman $-0.00$), so the shares are not meaningful.}
\label{tab:cov}
\begin{ruledtabular}
\begin{tabular}{lccc}
model & $4\zeta$ & $2\seig$ & $2\srot$ \\
\hline
si4096 ACE 2~K / 10~K & $-0.34$ / $-1.08$ & $+0.18$ / $+0.37$ & $+1.16$ / $+1.71$ \\
si4096 Keating 2~K / 10~K & $-0.09$ / $-0.31$ & $+0.29$ / $+0.07$ & $+0.80$ / $+1.24$ \\
R1 / R2 / R3, ACE 2~K & $-0.19$ / $-0.30$ / $-0.47$ & $+0.26$ / $+0.22$ / $+0.38$ & $+0.92$ / $+1.07$ / $+1.09$ \\
si512 ACE 2~K / 10~K & $-0.46$ / $-1.66$ & $+0.65$ / $+1.24$ & $+0.81$ / $+1.42$ \\
si512 GAP-18 2~K / 10~K$^a$ & $-0.94$ / --- & $+0.91$ / --- & $+1.03$ / --- \\
si4096 GAP-18 2~K / 10~K & $-0.68$ / $-2.25$ & $+0.20$ / $+0.44$ & $+1.48$ / $+2.81$ \\
\end{tabular}
\begin{tabular}{lccc}
 & si4096 ACE & si4096 Keating & \\
\hline
separation 0--1 / 1--3 / 3--10 / $>10\,\eta$ & 0.01 / 0.28 / 0.41 / 0.30 & 0.08 / 0.18 / 0.34 / 0.39 & \\
same, $\eta=10$~K & 0.32 / 0.43 / 0.27 / $-0.02$ & 0.24 / 0.33 / 0.34 / 0.09 & \\
pair mean frequency 0--10 / 10--30 / 30--45 / $>45$~meV & 0.39 / 0.54 / 0.09 / $-0.03$ & 0.23 / 0.69 / 0.06 / 0.01 & \\
same, $\eta=10$~K & 0.22 / 0.68 / 0.15 / $-0.06$ & 0.07 / 0.74 / 0.16 / 0.03 & \\
\end{tabular}
\end{ruledtabular}
\end{table}
In every separation bin up to $10\eta$ the binned rotation term is negative at high-$\zeta$ sites (top
quartile) and positive at low-$\zeta$ sites (bottom quartile).

What sets the width of this ``mixing window''? We compared several measures of how quickly correlations between pairs of modes decay with their separation in frequency (Table~\ref{tab:decay}).

\begin{table}[h]
\caption{Decay widths (separation at half maximum, meV) of pair correlations among the 10--30~meV modes of
si4096, and the half-point of the cumulative rotation covariance (30 logarithmic bins, 0.02--20~meV; the
half-points are interpolated in $\ln\omega$ within the bin where the cumulative share crosses one half;
pairs of 10--30~meV modes).}
\label{tab:decay}
\begin{ruledtabular}
\begin{tabular}{lcc}
 & ACE & Keating \\
\hline
co-localization $g$ & 2.3 & 2.6 \\
local overlap $h$ & 1.5 & 2.4 \\
$\langle|P_{ik}|^2\rangle$ & non-monotonic & 2.7 \\
$F$-weighted row overlap $r^2$, $\eta=2$ / 10~K & 1.3 / 1.7 & 3.7 / 3.5 \\
$N_{ik}^2$, $\eta=2$ / 10~K & 0.34 / no decay below 10~meV & 0.12 / 0.81 \\
cov$(\srot,\zeta)$ half-point, 4~\AA, $\eta=2$ / 10~K & 0.9 / 1.4 & 1.3 / 1.9 \\
cov$(\srot,\zeta)$ half-point, single sites, $\eta=2$ / 10~K & 1.4 / 2.0 & 1.6 / 2.4 \\
share of cov$(\srot,\zeta)$ from pairs with \emph{both} modes in 10--30~meV, 4~\AA, $\eta=2$ / 10~K & 0.43 / 0.51 & 0.59 / 0.64 \\
\end{tabular}
\end{ruledtabular}
\end{table}
The local overlap measures decay smoothly and identically at both broadenings. The current-correlation
measures either depend strongly on $\eta$ ($N^2$, which contains the AF weight) or differ between the
potentials without a common scale; they do not explain the window. The rebuilt $\srot$ used for this
analysis agrees with the sensitivity code to $10^{-4}$ (near-degenerate pairs omitted). Thus the window is set by where modes \emph{live}, not by how they carry current.

\paragraph*{Local return probability.} For the 10--30~meV band we computed the site-averaged return
probability of vibrational weight, $R(t)=\sum_{ik}H_{ik}\cos[(\omega_i-\omega_k)t]/\sum_{ik}H_{ik}$ with
$H_{ik}=\sum_{\ell}c_{ik}({\ell})^2$, which is the Fourier partner of $h$. Weight leaves its atom quickly: $R$ falls to
$1/e$ of its initial excess over $R(\infty)=0.0013$ at 0.146~ps (ACE) and 0.136~ps (Keating), i.e.
$\hbar/\tau=4.5$ and 4.8~meV, set by the band width rather than by the window. Between 0.2 and 2~ps,
$R-R(\infty)$ falls with local logarithmic slopes between $-2$ and $-4$ (ACE) and oscillates for Keating,
so the tail is not the $t^{-3/2}$ law of three-dimensional diffusion in this cell; the diffusion constants
implied by $R-R(\infty)=v_0(4\pi Dt)^{-3/2}$ (11--33~\AA$^2$/ps) are smaller than the AF values by factors of 3--9.
AF mode diffusivities (classical heat capacities, $\eta=2$~K), averaged over 10--30~meV: 94~\AA$^2$/ps
(0.94~mm$^2$/s, ACE) and 160~\AA$^2$/ps (1.60~mm$^2$/s, Keating); the classical $\kappa$ rebuilt from them,
1.331 and 1.862~\WmK{}, reproduces the direct values. We conclude that the time domain is not the place to see the window in a cell of this size; the frequency domain is.

\section{The 32\,768-atom replica}\label{sec:replica}
A larger cell supports longer-wavelength modes, and it is natural to ask whether the results of the main text survive them. Rather than build a new network, we used a replica of si4096, which adds the long waves while keeping the local structure fixed. The $2\times2\times2$ replica of si4096 has wave vectors $\bm k=\pi(n_1,n_2,n_3)/L$, $n_i\in\{0,1\}$, at
which the Bloch phases are $\pm1$, so each block $D(\bm k)=\sum_{\bm L}D(0,\bm L)e^{i\bm k\cdot\bm L}$ is a real $12\,288\times12\,288$ matrix
corresponding to 4096 atoms. The force constants are truncated at 11~\AA, with the acoustic sum rule restored;
no pair has two periodic images within this range, so the construction is exact. Because $\Gamma^\alpha$ is
translation invariant on the replica, $P^\alpha$ and $N_{ik}$ are block diagonal in $\bm k$, and every
quantity in Eqs.~(\textcolor{red}{1})--(\textcolor{red}{10}) of the main text is a sum over blocks. In particular,
$\kappa_{\rm rep}=\tfrac18\sum_{\bm k}\kappa_{\bm k}$. A block calculation changes atom ${\ell}$ in every one of its
eight images at once, a perturbation with the periodicity of the 4096-atom cell, which is why the problem
stays block diagonal: the derivative of $\kappa_{\rm rep}$ for the eight images together is
$\tfrac18\sum_{\bm k}\km^{(\bm k)}({\ell})$, and since the images are equivalent, the derivative for a single replica
atom is one eighth of that, $\tfrac1{64}\sum_{\bm k}\km^{(\bm k)}({\ell})$. The same holds for $\zeta$, while
$\dR({\ell})^2=\tfrac18\sum_{\bm k}\dR^{(\bm k)}({\ell})^2$ is a property of each image. The constant factors do not affect
any rank statistic, and the Euler sums are consistent: summing the single-atom derivatives over all
32\,768 atoms gives $\tfrac18\sum_{\bm k}\sum_{\ell}\km^{(\bm k)}({\ell})$. Each block was run
through the same code as a cell. The Euler rule closes to within $10^{-6}$~\WmK{} in all 24 block runs (8 blocks, 3 temperatures), and the
$\bm k=0$ block reproduces the 4096-atom cell ($\kappa=0.37834$ against 0.37831~\WmK{} at 50~K).

\begin{table}[h]
\caption{Replica against cell (ACE, $\eta=2$~K, 4~\AA\ averages; 95\% block-bootstrap intervals).}
\label{tab:replica}
\begin{ruledtabular}
\begin{tabular}{lcccccc}
 & \multicolumn{2}{c}{50~K} & \multicolumn{2}{c}{100~K} & \multicolumn{2}{c}{300~K} \\
 & cell & replica & cell & replica & cell & replica \\
\hline
$\kappa$ (\WmK{}) & 0.378 & 0.397 & 0.764 & 0.783 & 1.218 & 1.237 \\
lowest mode (meV) & 3.9 & 2.0 & 3.9 & 2.0 & 3.9 & 2.0 \\
$S(\km,\zeta)$ & $-0.48$ & $-0.71$ & $-0.53$ & $-0.72$ & $-0.41$ & $-0.56$ \\
$S(\km,\zeta\,|\,\dR)$ & $-0.23$ & $-0.49$ & $-0.17$ & $-0.38$ & $+0.02$ & $-0.05$ \\
 & $[-0.32,-0.15]$ & $[-0.55,-0.41]$ & $[-0.24,-0.10]$ & $[-0.45,-0.32]$ & $[-0.05,+0.09]$ & $[-0.12,+0.00]$ \\
$S(\zeta,\dR)$ & $-0.63$ & $-0.69$ & $-0.70$ & $-0.72$ & $-0.71$ & $-0.72$ \\
$S(\km,\dR)$ & $+0.51$ & $+0.63$ & $+0.62$ & $+0.77$ & $+0.61$ & $+0.75$ \\
\end{tabular}
\end{ruledtabular}
\end{table}

\paragraph*{The low-temperature residual (replica, 50~K).} Splitting $\km$ by restricting $F$ to pairs with
both modes below 10~meV (exact, since $\kappa$ and its gradient are linear in $F$; the masked runs close
the Euler rule in every block): the propagon-pair part carries $\kappa=0.215$ of 0.397~\WmK{} and 78\% of
cov$(\km,\zeta)$. Spearman with $\zeta$: $-0.59$ (propagon pairs), $-0.66$ (rest); partial at fixed
$\dR$: $-0.40$ $[-0.47,-0.32]$ and $-0.36$ $[-0.44,-0.29]$. Partial of the full $\km$ and $\zeta$ at fixed
$\dR$ averaged over 4, 6, 8 and 12~\AA: $-0.49$, $-0.56$, $-0.63$, $-0.68$; over the same radii
$S(\zeta,\dR)$ falls from $-0.69$ to $-0.29$. With the quantum amplitude
$\dR_{\ell}^2=\sum_ip_i({\ell})(\hbar/2m\omega_i)\coth(\hbar\omega_i/2k_BT)$ averaged over the eight $k$-blocks, the
partial of $\km$ and $\zeta$ is $-0.60$ $[-0.65,-0.55]$ at 50~K, $-0.44$ $[-0.50,-0.39]$ at 100~K and $-0.06$
$[-0.12,-0.01]$ at 300~K, against $-0.49$, $-0.38$ and $-0.05$ with the classical amplitude; the classical and quantum amplitudes produce
similar site rankings (Spearman 0.93, 0.99 and 1.00). By term (replica, 4~\AA): at 50~K the rotation term carries 85\% of
cov$(\km,\zeta)$ and keeps a partial correlation with $\zeta$ of $-0.49$ $[-0.56,-0.42]$ at fixed $\dR$; the
eigenvalue term carries 25\% with a partial of $-0.28$ $[-0.37,-0.19]$. At 300~K the corresponding values
are 108\% and $-0.10$ $[-0.16,-0.04]$ (rotation), 25\% and $-0.29$ $[-0.37,-0.22]$ (eigenvalue). Thus at room temperature local softness accounts for the anti-correlation, in the replica as in the cell; at low temperature the long waves add a component that it does not capture, and that component lives chiefly in the rotation term.

\paragraph*{Joint conditioning on spectral moments.} The low-$T$ residual of $S(\km,\zeta\,|\,\dR)$ in the replica (4~\AA) is
$-0.49$ at 50~K and $-0.38$ at 100~K. Adding the quantum amplitude as a second conditioner gives $-0.29$ $[-0.37,-0.21]$ and
$-0.19$ $[-0.26,-0.13]$. The two amplitudes are strongly collinear (rank correlation 0.97 at 50~K, 0.996 at 100~K; variance
inflation factors 15 and 130), so we test the drop against a null in which the second field is translated by a random
vector of the periodic cell, which keeps its distribution and spatial correlation but breaks its alignment with the maps:
200 such surrogates leave the partial at $-0.49$ $[-0.49,-0.48]$ and $-0.38$ $[-0.39,-0.37]$ ($p<0.005$). The local
moments $m_k(\ell)=\sum_np_{\ell n}\omega_n^k$ tell the same story: $m_{-2}$ is the classical amplitude (rank correlation 1.000),
and adding $m_{-1}$ gives $-0.30$ and $-0.23$; three and five moments give $-0.29$ and $-0.26$ at 50~K. Band weights below
10~meV recover less ($-0.41$) than band weights over the whole spectrum ($-0.28$). At 300~K the two amplitudes coincide
and nothing is gained. Thus the residual depends on the shape of the local spectrum, not only on one of its moments, and
a part ($-0.26$ with five moments at 50~K) is not captured by the local spectrum at all.

\section{Paracrystalline models}\label{sec:para}
\begin{figure}[h]
\includegraphics[width=\textwidth]{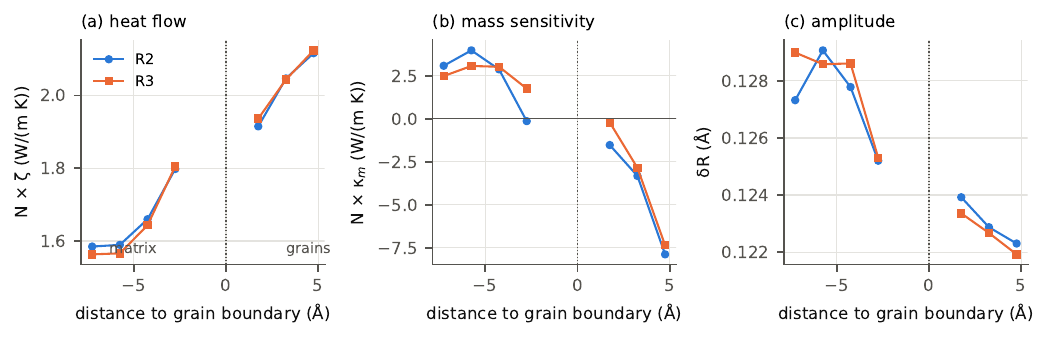}
\caption{R2 and R3 (ACE, $\eta=2$~K, single sites averaged in 1.5~\AA\ bins of the signed distance to the
grain boundary; negative: amorphous matrix, positive: grains). (a) Local heat flow $N\zeta$. (b) Mass
sensitivity $N\km$. (c) Classical vibrational amplitude $\dR$ (300~K).}
\label{fig:para}
\end{figure}

Paracrystalline silicon, with small grains embedded in an amorphous matrix, is a natural test of the picture: the grains are stiff and the matrix is soft. Atoms were classified with the ``identify diamond structure'' modifier of OVITO (version 3.16.1);
crystalline atoms are those in cubic or hexagonal diamond environments. The classification gives 100\% for
perfect diamond and 0.1\% (si4096) and 0.07\% (BM10k) for the continuous random networks, and 18.1\% (R2) and
24.3\% (R3), the fractions quoted in Ref.~\cite{Gautam2026lattice}. The signed distance to the grain
boundary is the distance from a crystalline atom to the nearest non-crystalline atom (positive), or minus
the distance from a non-crystalline atom to the nearest crystalline one; no atom lies within about a bond
length of the other region, hence the gap at zero in Fig.~\ref{fig:para}.

\paragraph*{Exact test of the grain-interior response.} Scaling the masses of all grain-interior atoms
($d>3$~\AA; 59 in R2, 97 in R3) by $f$ and recomputing $\kappa$ exactly ($\eta=2$ / 10~K), against the linear
prediction $-\sum s_m\ln f$ in parentheses: R2, $f=0.95$: $-0.47$ ($-0.33$) / $-0.18$ ($-0.09$)\%; $f=1.05$:
$+0.19$ ($+0.32$) / $-0.01$ ($+0.08$)\%; $f=1.2$: $-0.73$ ($+1.18$) / $-0.99$ ($+0.30$)\%. R3, $f=0.95$: $-0.72$
($-0.54$) / $-0.22$ ($-0.07$)\%; $f=1.05$: $+0.34$ ($+0.52$) / $-0.07$ ($+0.07$)\%; $f=1.2$: $-0.51$ ($+1.93$) /
$-1.66$ ($+0.26$)\%. The response is dominated by second-order terms beyond a few per cent, as for the crystal layer of the interface model. In other words, the linear map is a reliable guide to small changes of the grains, and only a rough one to large changes.

\begin{table}[h]
\caption{Crystalline fraction and conductivity; and $N\zeta$, $N\km$ (\WmK{}) and $\dR$ (\AA) by region.}
\label{tab:pararegion}
\begin{ruledtabular}
\begin{tabular}{lccccc}
model & crystalline & $\kappa$, $\eta=2$ / 10~K & region & $N\zeta$ / $N\km$ & $\dR$ \\
\hline
si4096, BM10k & 0.1\% & 1.22--1.23 / 0.95 & & & \\
R2 & 18.1\% & 1.72 / 1.28 & grain interior ($d>3$~\AA) & 2.05 / $-3.79$ & 0.1228 \\
   &        &             & matrix ($d<-3$~\AA) & 1.62 / $+3.14$ & 0.1281 \\
R3 & 24.3\% & 1.74 / 1.32 & grain interior & 2.06 / $-3.82$ & 0.1225 \\
   &        &             & matrix & 1.60 / $+3.05$ & 0.1286 \\
R1 & 92.7\% & 8.15 / 3.29 & grain interior & 8.42 / $+2.12$ & 0.1220 \\
   &        &             & matrix & 4.69 / $+108.5$ & 0.1487 \\
\end{tabular}
\end{ruledtabular}
\end{table}

\section{Design tests}\label{sec:design}
A sensitivity map is a first-order statement, and a designer will want to know how far it can be trusted for finite changes. Here are the details of the exact tests summarized in the main text. All tests use si4096 (ACE), exact $\kappa$ from one diagonalization per perturbed structure, $T=300$~K,
quantum heat capacities. Linear predictions: masses $-\sum_{\ell}s_m({\ell})\ln(m'/m)$; bonds
$\epsilon\sum s_b$ over bonds with both atoms in the region.

\paragraph*{Regions.} We changed compact regions around five centres per class, with radii of 2.6, 4 and 6~\AA\ (about 5, 14 and 46 atoms). The ratio of the exact to the linear change for regions centred on the most sensitive sites is: $^{30}$Si 1.10, 1.10, 1.21; bonds
$\pm5\%$ (4~\AA) 0.93 and 1.06; bonds $\pm20\%$ 0.67 and 1.17; Ge ($\eta=2$~K) 2.3, 3.0, 4.0. For Ge at
$\eta=10$~K, $\Delta\kappa$ for top / random / bottom centres is $-0.41/-0.42/-0.40\%$ (5 atoms),
$-0.98/-0.99/-0.93\%$ (14 atoms) and $-2.33/-2.38/-2.42\%$ (46 atoms). Thus isotopes and modest bond changes are nearly linear, while a compact cluster of heavy atoms is not, and does not care where it sits.

\paragraph*{Pairs.} For pairs of Ge atoms we define the interaction as the exact change for the pair minus the sum of the two single changes, relative to that sum. si4096: 6.5--8.7\% at 2.35~\AA, 1.5--1.9\% at 5~\AA, 0.5--2.0\% at 10~\AA\ (45 evaluations).
In si512 (600 evaluations) the interaction is nearly independent of distance at $\eta=2$~K, a small-cell effect. So substituents more than a few \AA\ apart act nearly independently, and Table~\ref{tab:ge} shows what the map can do with that.

\begin{table}[h]
\caption{Dispersed Ge (sites at least 5~\AA\ apart): $\Delta\kappa$ (\%). Random: mean $\pm$ standard
deviation over five sets. Last column: random sets drawn without the separation constraint.}
\label{tab:ge}
\begin{ruledtabular}
\begin{tabular}{lccccc}
$\eta$, fraction & most sensitive & random & least sensitive & (top$-$random)/sd & random, unconstrained \\
\hline
2~K, 1\% & $-7.96$ & $-4.64\pm0.25$ & $-3.36$ & 13 & $-4.76\pm0.19$ \\
2~K, 3\% & $-16.82$ & $-11.84\pm0.31$ & $-9.10$ & 16 & $-12.42\pm0.18$ \\
2~K, 5\% & $-22.36$ & $-17.15\pm0.23$ & $-14.09$ & 23 & $-18.69\pm0.17$ \\
10~K, 1\% & $-5.12$ & $-3.41\pm0.11$ & $-3.04$ & 15 & $-3.39\pm0.11$ \\
10~K, 3\% & $-12.21$ & $-9.23\pm0.21$ & $-8.34$ & 14 & $-9.26\pm0.26$ \\
10~K, 5\% & $-17.17$ & $-14.11\pm0.16$ & $-13.04$ & 19 & $-14.27\pm0.13$ \\
\end{tabular}
\end{ruledtabular}
\end{table}

\section{Interface}\label{sec:interface}
The interface model has a crystalline layer between two amorphous regions, and one may wonder whether it has vibrations of its own. It does not, in any strong sense. The fraction of modes with 20--80\% of their weight in the crystalline layer ($z<18$~\AA, 32\% of the atoms),
and in parentheses the mean weight in the crystal: below 10~meV 0.42 (0.19); 10--30~meV 0.89 (0.32);
30--45~meV 0.73 (0.27); 45--56~meV 0.69 (0.24); above 56~meV 0.84 (0.42). The mean weight in quasi-localized soft modes (modes below 12.4~meV in the lowest 10\% of
participation ratio in that band) is 0.004 in the crystal, 0.005 within 4~\AA\ of the interfaces and 0.011 in the
amorphous region; the weight below 12.4~meV is 0.048, 0.060 and 0.104. Within regions (single sites, $\eta=10$~K;
interfaces: atoms within 4~\AA\ of $z=0$ or $z=18$~\AA), $S(\zeta,\dR)=-0.54$, $-0.88$, $-0.54$ and
$S(\km,\dR)=+0.44$, $+0.40$, $+0.32$ (amorphous, interfaces, crystal). Thus the soft modes live on the amorphous side, and within each region the rattler statistics of the bulk reappear.

\section{Fully anharmonic check: per-atom Green--Kubo}\label{sec:gk}
Everything in the main text is harmonic. A fully anharmonic check is possible in principle, with Green--Kubo molecular dynamics and a per-atom heat current, but it is noisy and expensive; we did it for the 512-atom model only, and read it qualitatively.

\paragraph*{Method.} The per-atom heat current is $\bm j_i=(e_i^{\rm kin}+e_i^{\rm pot})\bm v_i-\bm S_i\cdot\bm v_i$,
with the centroid form of the many-body virial~\cite{Fan2015force}. We added centroid-stress support to the
\texttt{pair\_pace} style of LAMMPS (CPU version), accumulating $(\bm r_j-\bm r_i)\otimes(-\bm f_{ij})$ on the
atom that owns the energy; the patch agrees with finite differences to about $10^{-7}$, and $\sum_i\bm j_i$
equals the total heat flux of LAMMPS exactly. For si512 we ran four independent NVE trajectories of 1~ns
(time step 2~fs, after 20~ps NVT) at 300~K and at 125~K. The local map
$\zeta_{\rm GK}({\ell})=\langle\bm j_{\ell}(t)\cdot\bm B(t)\rangle/(3Vk_BT^2)$ with $\bm B(t)=\int_{t-\tau}^t\bm J$
sums to $\kappa_{\rm GK}(\tau)$ exactly. The harmonic reference uses classical heat capacities in the ACE
site-energy gauge.

\begin{table}[h]
\caption{Green--Kubo results for si512 (ACE; mean $\pm$ standard error over four trajectories).}
\label{tab:gk}
\begin{ruledtabular}
\begin{tabular}{lcc}
 & 300~K & 125~K \\
\hline
$\kappa_{\rm GK}$ at the plateau (\WmK{}) & $1.57\pm0.09$ ($\tau=10$~ps) & $1.67\pm0.07$ ($\tau=20$~ps) \\
harmonic AF, classical $C$, $\eta=2$~K & 1.21 & 1.21 \\
$\kappa_{\rm GK}$ at $\tau=20$~ps & $1.06\pm0.15$ & $1.67\pm0.07$ \\
map reproducibility, 4~\AA\ (split-half Pearson) & 0.84--0.98 & 0.86--0.97 \\
MD map vs harmonic $\zeta$, 4~\AA: $\tau=1$--2~ps / plateau & $-0.43$ to $-0.52$ / $+0.45$ & $-0.41$ to $-0.50$ / $+0.45$ \\
MD map vs $\dR$: $\tau=1$--2~ps / plateau & $+0.26$ to $+0.30$ / $-0.29$ & $+0.23$ to $+0.25$ / $-0.26$ \\
$\dR$ from MD vs harmonic: Pearson, slope & 0.83, 1.57 & 0.87, 1.39 \\
\end{tabular}
\end{ruledtabular}
\end{table}

The plateau values of $\kappa_{\rm GK}$ exceed the harmonic AF value of the same cell. We do not read
this as an anharmonic enhancement: the plateau was chosen by eye in a noisy integral, as the first broad maximum of the mean over four
trajectories (10~ps at 300~K, 20~ps at 125~K), and Table~\ref{tab:gk} gives the maps at short times and at the plateau
so that their dependence on the window can be judged; we have not attempted an objective criterion (at 300~K the integral falls to
$1.06\pm0.15$ at 20~ps), and in a 512-atom periodic cell the few low-frequency modes are long lived in
molecular dynamics but are represented in the AF value only through the broadening $\eta$. We use the
Green--Kubo runs only for the map comparisons below, not for the magnitude of $\kappa$.

At short integration times the local Green--Kubo map follows current fluctuations, which are large at the
rattlers; transport information appears near the plateau, later at lower $T$. At 300~K the integral turns
down beyond 10~ps because of a noisy tail ($\kappa_{\rm GK}(20\,{\rm ps})=1.06\pm0.15$). The comparison is
qualitative: the 512-atom system and 1~ns trajectories provide limited spatial and temporal sampling for a converged local map. Still, the change of sign of the map between short times and the plateau is just what the rattler picture leads one to expect: rattlers dominate the fluctuations of the current, not its transport.

\section{Bonds}\label{sec:bonds}
The main text concentrates on masses. The bond map is its companion, and here we ask what kind of bond motion it responds to. Each nearest-neighbour block $D_{{\ell}{\ell'}}$ is split into its central part $(nn^{\mathsf T})D_{{\ell}{\ell'}}(nn^{\mathsf T})$
(stretching) and the remainder (bending and non-central terms), with the same symmetric self-block
compensation as $s_b$; the two parts add to $s_b$ to $10^{-15}$.
\begin{table}[h]
\caption{Bond sensitivities: rank correlation with bond length, its stretching part, the bending share of the summed sensitivity, and the correlation of site stiffness with $\zeta$.}
\label{tab:bonds}
\begin{ruledtabular}
\begin{tabular}{lcccccc}
 & ACE si512 2~K & ACE si512 10~K & GAP si512 2~K & GAP si512 10~K & ACE si4096 2/10~K & Keating si4096 2~K \\
\hline
Spearman$(s_b,\text{length})$ & $+0.07$ & $+0.45$ & $+0.16$ & $+0.43$ & $+0.30$ / $+0.47$ & $-0.25$ \\
stretching part vs length & $+0.08$ & $+0.48$ & $+0.14$ & $+0.44$ & $+0.36$ / $+0.52$ & $-0.17$ \\
bending share of $\sum s_b$ & 79\% & 64\% & 69\% & 52\% & 71\% / 56\% & 85\% \\
site stiffness vs $\zeta$ (Pearson) & $+0.12$ & $+0.38$ & $+0.05$ & $+0.29$ & $+0.18$ / $+0.38$ & $+0.09$ \\
\end{tabular}
\end{ruledtabular}
\end{table}
Bending dominates the total bond sensitivity for all potentials. For the two machine-learned potentials the
length dependence is driven by stretching of long bonds; Keating reverses it (Sec.~\ref{sec:potentials}). Stiffening the
bonds of heat-carrying atoms raises $\kappa$ for all three potentials. Thus, as one would guess for a tetrahedral network, the angles matter more than the bond lengths.

\bibliography{refs}